\input harvmac 
\input epsf.tex

\overfullrule=0mm
\newcount\figno
\figno=0
\def\fig#1#2#3{
\par\begingroup\parindent=0pt\leftskip=1cm\rightskip=1cm\parindent=0pt
\baselineskip=11pt
\global\advance\figno by 1
\midinsert
\epsfxsize=#3
\centerline{\epsfbox{#2}}
\vskip 15pt
{\bf Fig. \the\figno:} #1\par
\endinsert\endgroup\par
}
\def\figlabel#1{\xdef#1{\the\figno}}
\def\encadremath#1{\vbox{\hrule\hbox{\vrule\kern8pt\vbox{\kern8pt
\hbox{$\displaystyle #1$}\kern8pt}
\kern8pt\vrule}\hrule}}
\def\encadre#1{\vbox{\hrule\hbox{\vrule\kern8pt\vbox{\kern8pt#1\kern8pt}
\kern8pt\vrule}\hrule}}

\def\tvi{\vrule height 10pt depth 6pt width 0pt}
\def\tv{\tvi\vrule}

\def\IR{\relax{\rm I\kern-.18em R}}
\font\cmss=cmss10 \font\cmsss=cmss10 at 7pt
\def\IZ{\relax\ifmmode\mathchoice
{\hbox{\cmss Z\kern-.4em Z}}{\hbox{\cmss Z\kern-.4em Z}}
{\lower.9pt\hbox{\cmsss Z\kern-.4em Z}}
{\lower1.2pt\hbox{\cmsss Z\kern-.4em Z}}\else{\cmss Z\kern-.4em Z}\fi}
\def\buildrel#1\under#2{\mathrel{\mathop{\kern0pt #2}\limits_{#1}}}

\Title{SPhT/96-062}
{{\vbox {
\bigskip
\centerline{Meanders: A Direct Enumeration Approach}
}}}
\bigskip
\centerline{P. Di Francesco,}
\medskip
\centerline{O. Golinelli}
\medskip
\centerline{and} 
\medskip
\centerline{E. Guitter\footnote*{e-mails:
philippe,golinel,guitter@spht.saclay.cea.fr},}

\bigskip

\centerline{ \it Service de Physique Th\'eorique, C.E.A. Saclay,}
\centerline{ \it F-91191 Gif sur Yvette, France}

\vskip .5in
We study the statistics of semi-meanders, i.e. configurations of a set
of roads crossing a river through $n$ bridges, and possibly winding
around its source, as a toy model for compact folding of polymers.  By
analyzing the results of a direct enumeration up to $n=29$, we perform
on the one hand a large $n$ extrapolation and on the other hand we
reformulate the available data into a large $q$ expansion, where $q$ is
a weight attached to each road.  We predict a transition at $q=2$
between a low-$q$ regime with irrelevant winding, and a large-$q$
regime with relevant winding.

\noindent
\Date{06/96}

\nref\HMRT{K. Hoffman, K. Mehlhorn, P. Rosenstiehl and R. Tarjan, {\it
Sorting Jordan sequences in linear time using level-linked search
trees}, Information and Control {\bf 68} (1986) 170-184.}
\nref\ARNO{V. Arnold, {\it The branched covering of $CP_2 \to S_4$,
hyperbolicity and projective topology},
Siberian Math. Jour. {\bf 29} (1988) 717-726.}
\nref\KOSMO{K.H. Ko, L. Smolinsky, {\it A combinatorial matrix in
$3$-manifold theory}, Pacific. J. Math {\bf 149} (1991) 319-336.}
\nref\TOU{J. Touchard, {\it Contributions \`a l'\'etude du probl\`eme
des timbres poste}, Canad. J. Math. {\bf 2} (1950) 385-398.}
\nref\LUN{W. Lunnon, {\it A map--folding problem},
Math. of Computation {\bf 22}
(1968) 193-199.}
\nref\LZ{S. Lando and A. Zvonkin, {\it Plane and Projective Meanders},
Theor. Comp.  Science {\bf 117} (1993) 227-241, and {\it Meanders},
Selecta Math. Sov. {\bf 11} (1992) 117-144.}
\nref\DGG{P. Di Francesco, O. Golinelli and E. Guitter, {\it Meander,
folding and arch statistics}, to appear in Journal of Mathematical and
Computer Modelling (1996).}
\nref\MAK{Y. Makeenko, {\it Strings, Matrix Models and Meanders}, proceedings
of the 29th Inter. Ahrenshoop Symp., Germany (1995); 
Y. Makeenko and H. Win Pe, {\it Supersymmetric matrix models and the meander
problem}, preprint ITEP-TH-13$/$95 (1996); G. Semenoff and
R. Szabo {\it Fermionic Matrix Models} preprint UBC$/S96/2$ (1996).}
\nref\NTLA{P. Di Francesco, O. Golinelli and E. Guitter, {\it
Meanders and the Temperley-Lieb algebra}, Saclay preprint T96/008 (1996).}
\nref\SLO{N. Sloane, {\it The on--line encyclopedia of integer sequences},
\hfill\break e-mail: sequences@research.att.com}
\nref\BAX{R. Baxter, {\it Exactly solved models in statistical mechanics},
Academic Press, London (1982).}


\newsec{Introduction}

The meander problem is a simply stated combinatorial question:
count the number of configurations of a closed non-self-intersecting road 
crossing an infinite river through a given number of bridges.
Despite its apparent simplicity, this problem still awaits 
a solution, if only for asymptotics when the number of bridges 
is large.
The problem emerged in various contexts ranging from mathematics
to computer science \HMRT. In particular, Arnold re-actualized it
in connection with Hilbert's 16th problem,
namely the enumeration of ovals of planar algebraic curves \ARNO,
and it also appears in the classification of 3-manifolds \KOSMO.

Remarkably, the meander problem can be rephrased in the physical language
of critical phenomena,
through its equivalence with a particular problem of 
Self-Avoiding Walks:
the counting of the compact foldings of a linear chain.

Several techniques have been applied to this problem:
direct combinatorial approaches \TOU\ \LUN,
random matrix model techniques \LZ\ \DGG\ \MAK, an
algebraic approach using
the Temperley-Lieb algebra and Restricted Solid-On-Solid models
\NTLA. Several exact results have been obtain on the way for 
meander-related issues, including exact sum rules for
meandric numbers \DGG, the solution of the somewhat simpler irreducible
meander problem \LZ\ \DGG, and the calculation of a
meander-related determinant \NTLA\ \KOSMO. 

The present paper is dedicated to a more direct {\it enumerative}
approach and a thorough analysis of its results in the spirit of
critical phenomena.  The meander problem is generalized to include the
case of several non-intersecting but possibly interlocking roads with a
weight $q$ per road.  The corresponding generating functions are
analyzed as functions of $q$. In particular, we derive their large $q$
asymptotic expansion in powers of $1/q$.

The paper is organized as follows. 
In sect.2, we give the basic definitions
of meanders and semi-meanders (which correspond 
to the same problem with a semi-infinite river with a source,
around which the roads are free to wind), as well as some 
associated observables such as the winding.
We further give exact solutions to the meander and semi-meander problems
at two particular 
values of $q$: $q=1$, where they reduce to a
random walk problem, and $q=\infty$, dominated by simple
configurations.  In sect.3, we explain how to enumerate the
semi-meanders for arbitrary number $n$ of bridges, using a 
fundamental recursive construction. 
After implementation
on a computer, this procedure allowed us to find the
semi-meander numbers with up to $n=29$ bridges.
These data are presented and then analyzed by a direct large 
$n$ extrapolation.  On the way we also  
confirm the scaling hypotheses borrowed from the theory of critical
phenomena.  Evidence is found for a phase transition for semi-meanders
at a value of $q=q_c\simeq 2$ between a low-$q$ and a large-$q$
regimes, discriminated by the relevance of winding around the source.
In sect.4, we show how to use the above data to generate a large-$q$
expansion for most of the interesting quantities.  This expansion
provides an accurate description of the whole $q>q_c$ phase.  In
sect.5, we analyze the break-down of this expansion, which gives rise
to the $q<q_c$ phase. Sect.6 briefly describes the small-$q$ expansion
of the problem. We gather our conclusions in sect.7.  The more
technical details are left in appendices.

\newsec{The meander problem}

\subsec{Definitions, observables}

A {\it meander} of order $n$ is a planar configuration of a
non-self-intersecting loop (road) crossing a line (river), through a
given number $2n$ of points (bridges).  We consider as equivalent any
two configurations which may be continuously deformed into each other,
keeping the river fixed (this is therefore a topological equivalence). 
The number of inequivalent meanders of order $n$ is denoted by
$M_n$. For instance, we have $M_1=1$, $M_2=2$, $M_3=8$... 
More numbers can be found in \LZ\ \DGG\ \SLO.
 
We stumbled on the meander problem by trying to enumerate the distinct
{\it compact folding} configurations of a closed polymer, i.e. the
different ways of folding a closed chain of $2n$ identical constituents
onto itself.  The best image of such a closed polymer is that of a
closed strip of $2n$ identical stamps, attached by their edges, serving
as hinges in the folding process: a compactly folded configuration of
the strip is simply a folded state in which all the stamps are piled up
on top of one of them.

\fig{The mapping between compactly folded closed strip of stamps and meanders.
We display a compact folding configuration (a) of a closed strip with $2n=6$
stamps. To transform it into a meander, first draw a (dotted) line
through the centers of the stamps and close it to the left of the picture.
Then cut the bottom right hinge (empty circle) and pull its ends apart
as indicated by the arrows,
so as to form a straight line (b): 
the straight line forms the river, 
and the dashed line
the road of the resulting meander.}{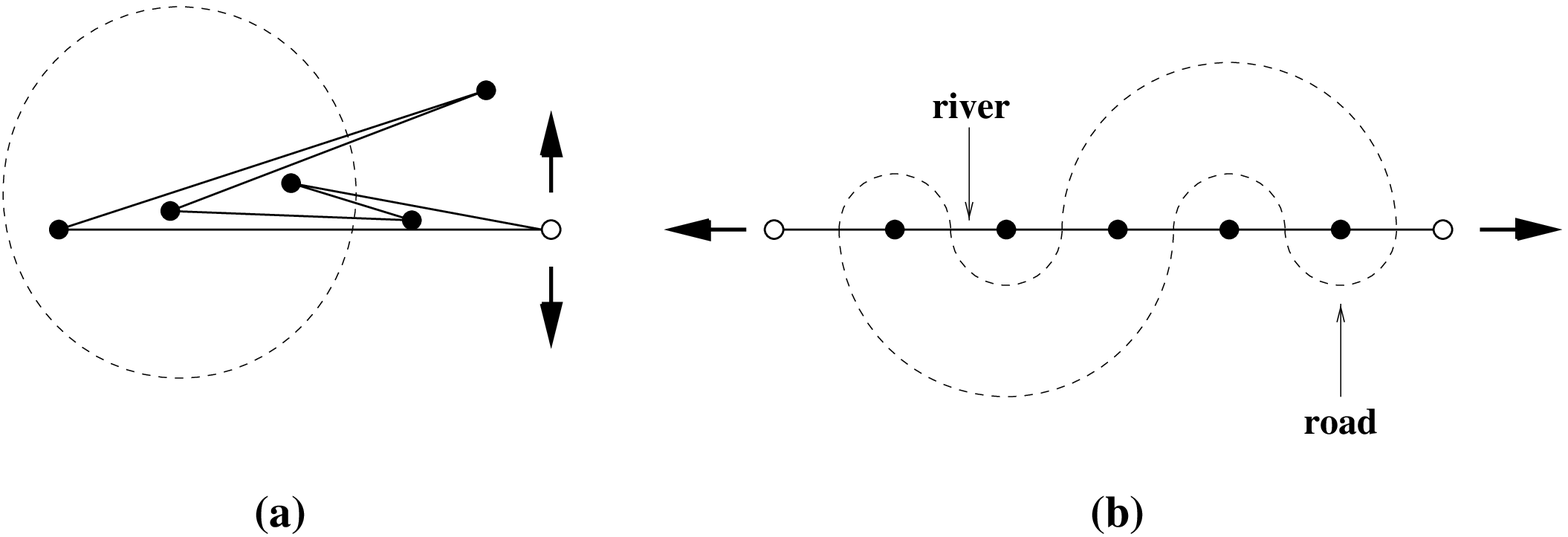}{10.5 cm}
\figlabel\compac

Such a compactly folded configuration is easily identified with a
meander configuration as depicted in Fig.\compac.  Draw a closed line
(road) passing though the centers (bridges) of all the piled-up
monomers, then open one hinge of the polymer (we choose to always open
the bottom right one) and pull the stamps apart so as to form a
straight line: the latter is identified with the river, whereas the
distorted line becomes the road of the resulting meander.

\fig{The $4$ inequivalent foldings of a strip of $3$ stamps. The fixed
stamp is indicated by the empty circle: it is attached to a support
(shaded area). The other circles correspond to the edges of the
stamps.}{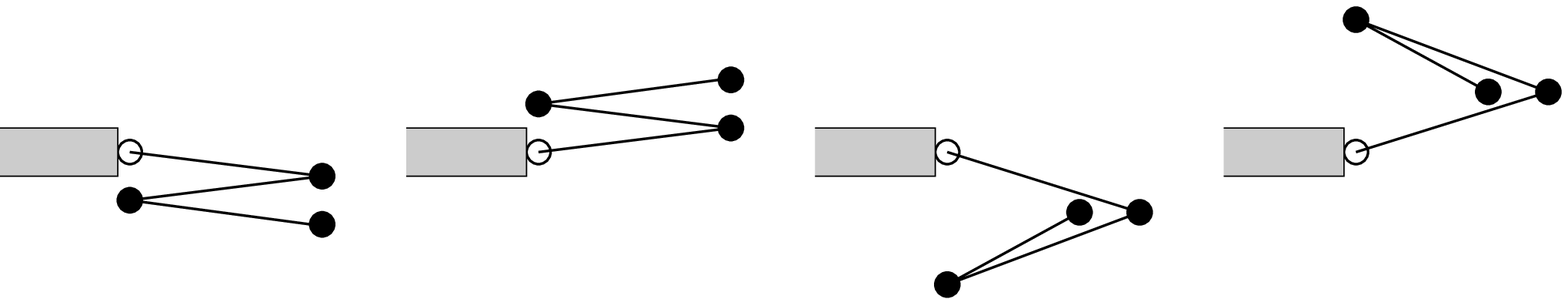}{9.5truecm} \figlabel\stamp

When the strip of stamps is open (see Fig.\stamp), we decide to attach
the first stamp to a support,
preventing the strip from winding around it, while the last stamp has a 
free extremal edge.
In this case, a slightly generalized transformation maps any compactly
folded open configuration of $(n-1)$ stamps to what we will call a 
{\it semi-meander}
configuration of order $n$, in the
following manner. 

\fig{The mapping of a compactly folded configuration of $4$ stamps onto
a semi-meander of order $5$. (a) draw a (dashed) curve through the pile 
of stamps and the (shaded) support. (b) pull the free edge of the 
last stamp to form a half-line (the river with a source). (c) the result
is a semi-meander configuration of order $5$, namely that of a road, crossing
a semi-infinite river through $5$ bridges (the source of the river, 
around which the road is free to wind, is indicated by an 
asterisk).}{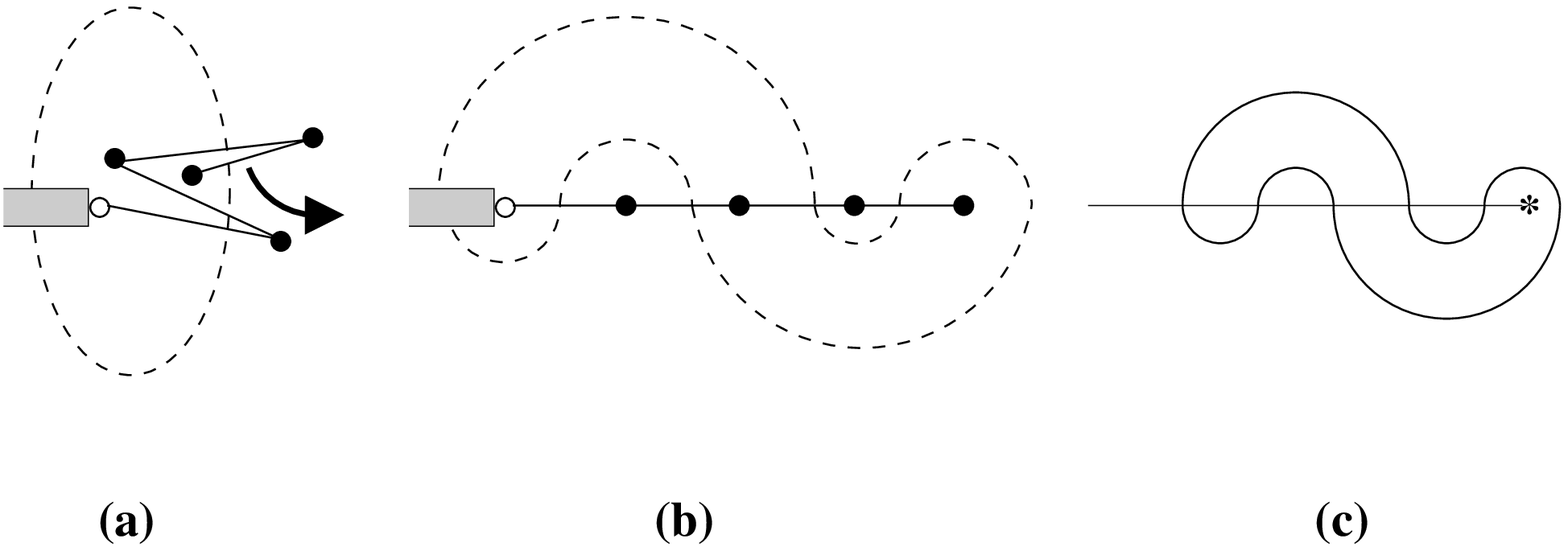}{10.cm}
\figlabel\transfoldsm

As shown in Fig.\transfoldsm, draw a curve (road) though the $(n-1)$
centers (bridges) of all the piled-up stamps, then close this curve
across the support (this last intersection is the $n$-th bridge), and
pull the free edge of the last stamp in order to form a straight
half-line (river with a source). The resulting picture is a
configuration of a road (the curve) crossing a semi-infinite river
(stamps and support) through $n$ bridges: this is called a semi-meander
configuration of order $n$. Note that the road in a semi-meander may
wind freely around the source of the river, and that consequently the
number of bridges may be indifferently even or odd, as opposed to
meanders.  The number of distinct semi-meanders of order $n$ is denoted
by ${\bar M}_n$. For instance, we have ${\bar M}_1=1$, ${\bar M}_2=1$,
${\bar M}_3=2$, ${\bar M}_4=4$... More numbers can be found in
\TOU\ \DGG\ and in appendix A.

Through its compact folding formulation, the semi-meander problem
is a particular reduction of the two-dimensional self-avoiding walk 
problem, in which only topological constraints are retained.
It is therefore natural to define, by analogy with self-avoiding walks
the connectivity $\bar R$ per stamp and the configuration exponent $\gamma$
which determine the large $n$ behavior of the semi-meander numbers as 
follows\foot{That the semi-meander numbers ${\bar M}_n$ actually have 
these leading asymptotics may be proved by deriving upper and lower 
bounds on $\bar R$.
See \DGG\ for further details.} 
\eqn\actiga{ {\bar M}_n ~ \sim ~ {\bar c}~{{\bar R}^n \over n^\gamma} }
The connectivity $\bar R$ may be interpreted as the average number
of possibilities of adding one stamp to the folded configurations.
The exponent $\gamma$ is characteristic of the (open) boundary condition 
on the strip of stamps.

\fig{The ``end-to-end distance" of the folded strip of stamps (a) is the
number ($w=1$ here) of stamps to be added to the strip (the added stamp 
is represented in dashed line), so that the new free end (empty circle) is
in contact with the infinity to the right. This coincides with the ``winding'' 
of the corresponding semi-meander (b), namely the number of bridges to be
added if we continue the river to the right of its source 
(dashed line).}{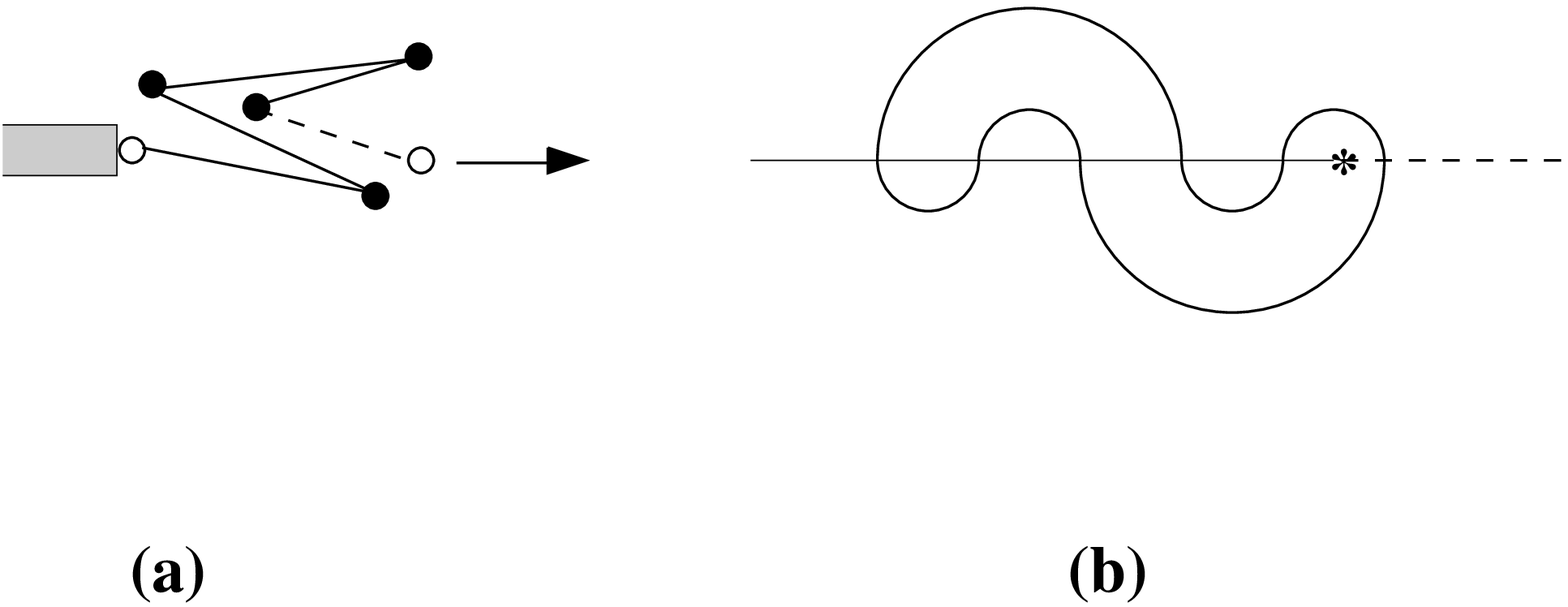}{8.cm}
\figlabel\wind

A natural observable for self-avoiding walks is the end-to-end distance. 
The corresponding notion for a compactly folded open strip of stamps is
the ``distance'' between the free end of the strip and, say the
support. This distance should also indicate how far the end of the
strip is buried inside the folded configuration.  It is defined as the
minimal length $w$ of a strip of stamps to be attached to the free end,
such that a resulting folding with $n-1+w$ stamps has its free end
outside of the folding, namely can be connected to the infinity to the
right of the folding by a half-line which does not intersect any
stamp.  Indeed, the infinity to the right can be viewed as the nearest
topological neighbor of the support, hence $w$ measures a distance from
the free end of the strip to the support.  This is illustrated in
Fig.\wind (a), with $n=5$ and $w=1$.  In the semi-meander formulation
(see Fig.\wind (b)), this distance $w$ is simply the {\it winding} of
the road around the source of the river, namely the number of bridges
to be added if we continue the river to the right of its source.  By
analogy with self-avoiding walks, we expect the average winding over
all the semi-meanders of order $n$ to have the leading behavior
\eqn\avwind{ \langle w \rangle_n~\equiv~ {1 \over {\bar M}_n}\sum_{\rm
semi-meanders} w~ \sim ~ n^\nu } where $\nu$ is some positive
(end-to-end) exponent $0\leq \nu \leq 1$, as $w$ is always smaller or
equal to $n$.

In this language, a meander of order $n$ is simply a semi-meander of
order $2n$ with winding $w=0$.  By analogy with closed (as compared to
open) self-avoiding walks, we expect the asymptotics \eqn\meas{ M_n ~
\sim ~ c ~ {R^{2n} \over n^\alpha} } where the connectivity per bridge
$R$ is the same as that for semi-meanders \actiga, $R=\bar R$, and the
configuration exponent $\alpha\neq \gamma$ is characteristic of the
closed boundary condition on the strip of stamps.

In the following, we will mainly focus our study on the semi-meander
numbers.

\subsec{Arches and connected components}

\fig{A semi-meander viewed as a particular meander:
the semi-infinite river must be opened up as indicated by the arrows.
This doubles the number of bridges in the resulting meander,
hence the order is conserved ($n=5$ here).
By construction,
the lower arch configuration of the meander is always a rainbow
arch configuration of same order. 
}{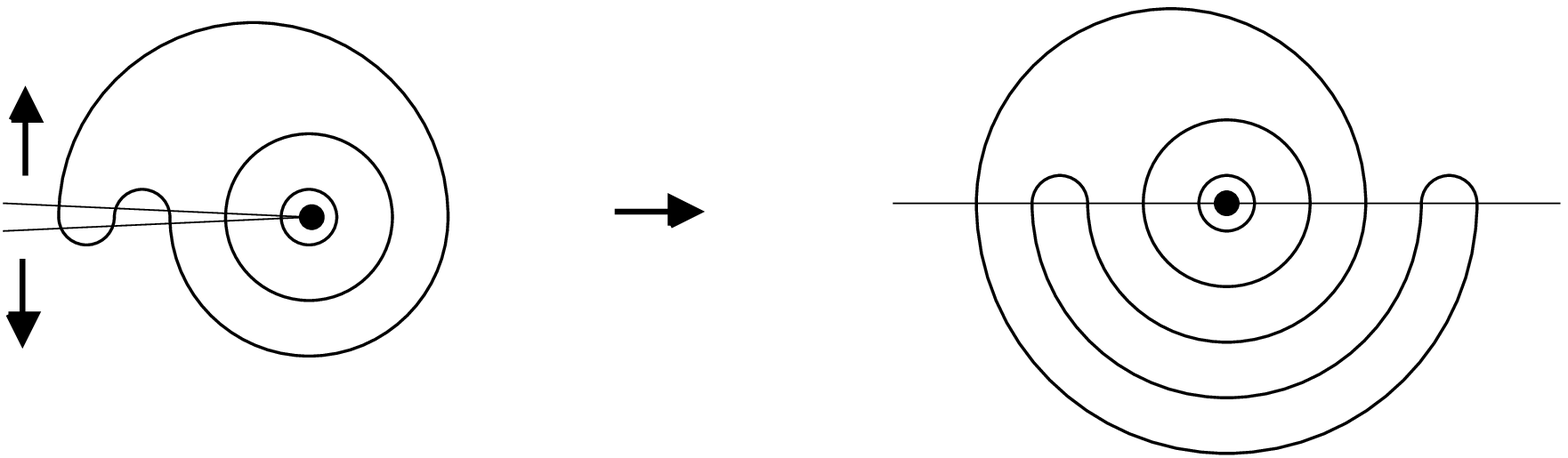}{10.cm}
\figlabel\semimean

Any semi-meander may be viewed as a particular meander by opening the
semi-infinite river as indicated by the arrows on Fig.\semimean. In the
process, the number of bridges is doubled, hence the order is
conserved.  The resulting meander however is very peculiar.  Note that
in general a meander is made of an upper (resp. lower) configuration
consisting of non-intersecting arches (arcs of road) connecting the
bridges by pairs above (resp. below) the river.  In the present case
the lower configuration is fixed: it is called the rainbow arch
configuration of order $n$ (the bridge $i$ is connected to the bridge
$(2n-i+1)$, $i=1,2,...,n$).  On the other hand, the upper arch
configuration may take any of the ${\bar M}_n$ values leading to
semi-meanders of order $n$.

There are however 
\eqn\catal{ c_n~=~{(2n)!\over n!(n+1)!} }
distinct arch configurations of order $n$ \DGG, as is readily proved by
recursion ($c_{n+1}=\sum_{0 \leq j \leq n} c_j c_{n-j}$, with $c_0=1$,
hence $c_1=1$, $c_2=2$, $c_3=5$, $c_4=14$,...:
the $c_n$ are called the Catalan numbers). 
Hence not all upper arch configurations, once supplemented by a lower
rainbow arch configuration of same order, lead to an opened
semi-meander (${\bar M}_n<c_n$).
This is because, in general, the corresponding object will have $k \geq
1$ connected components:  we call it a semi-meander of order $n$ with
$k$ connected components.  Indeed, if the river is folded back into a
semi-infinite one, we are simply left with a collection of $k$ possibly
interlocking semi-meanders of respective orders $n_1$, $n_2$,...,
$n_k$, with $n_1+n_2+...+n_k=n$.  We always have $1 \leq k \leq n$, and
$k=n$ only for the superposition of an upper and a lower rainbow
configurations, leading to $n$ concentric circles.  We denote by ${\bar
M}_n^{(k)}$ the number of inequivalent semi-meanders of order $n$ with
$k$ connected components. In particular, we have ${\bar
M}_n^{(1)}={\bar M}_n$ and ${\bar M}_n^{(n)}=1$ for all $n$.

The direct numerical study of the asymptotics of the numbers
${\bar M}_n^{(k)}$ turns out to be delicate, as the natural
scaling variable of the problem is the ratio $x=k/n$, which depends on $n$
and takes only a discrete set of values.
To circumvent this problem, we will study the generating function 
${\bar m}_n(q)$ for these numbers, also referred to as 
the {\it semi-meander polynomial}.
\eqn\semidef{ {\bar m}_n(q)~=~ \sum_{k=1}^n q^k ~{\bar M}_n^{(k)} }
This quantity makes it possible to study the large $n$ asymptotics
of the ${\bar M}_n^{(k)}$ in a global way, by use of extrapolation
techniques for all real values of $q$.
The semi-meander polynomial \semidef\ may be viewed as the partition 
function of a statistical assembly of
multicomponent semi-meanders of given order $n$, with a fugacity $q$
per connected component.  As such, it is expected to have an extensive large
$n$ behavior, namely
\eqn\asympto{ {\bar m}_n(q)~\sim ~ {\bar c}(q)~
{{\bar R}(q)^n \over n^{\gamma(q)}} }
where ${\bar R}(q)$ is the partition function per bridge, $\gamma(q)$ is 
a possibly varying exponent and ${\bar c}(q)$ a function independent of $n$. 
For $q \to 0$ ($k=1$),
we must recover the connected semi-meanders, namely that 
${\bar m}_n(q)/q \to {\bar M}_n$, i.e.
\eqn\recoq{ {\bar  R}(q) \to R \qquad \gamma(q) \to \gamma  
\qquad {\bar c}(q)/q \to {\bar c} }
(c.f. \actiga).
The notion of winding is well-defined for multi-component semi-meanders
as well, as the sum of the individual windings of each connected component,
namely the {\it total} number of times the various roads forming the 
semi-meander wind around the source of the river. Therefore we define
\eqn\defwik{ \langle w \rangle_n(q)~=~{1 \over {\bar m}_n(q)}
\sum_{{\rm multicomp.} \atop {\rm semi-meanders}} w ~q^k \sim n^{\nu(q)} }
where $\nu(q)$ is the generalized winding exponent for multi-component 
semi-meanders, satisfying $0 \leq \nu(q)\leq 1$.

Analogously, we define multi-component meanders of order $n$, 
as configurations
of $k$ non-intersecting roads ($1 \leq k \leq n$)
crossing the river through a total of $2n$
bridges, and denote by $M_n^{(k)}$ their number.
We also define the {\it meander polynomial}
\eqn\defmpo{ m_n(q)~=~ \sum_{k=1}^n q^k~ M_n^{(k)}  }
This is nothing but the restriction of \semidef\ with $n \to 2n$, to
semi-meanders with zero winding $w=0$.  We therefore expect the
asymptotics for large $n$
\eqn\asymean{ m_n(q) ~\sim ~ c(q) 
{ R(q)^{2n} \over n^{\alpha(q)}} }
In this estimate, the partition function per bridge ${R}(q)$ is expected
to be identical to that of semi-meanders ${\bar R}(q)$ only if 
the winding is irrelevant, namely if $\nu(q)$ is strictly less than $1$
\eqn\equalir{ {R}(q) ~=~ {\bar R}(q) \quad {\rm iff} \quad \nu(q)<1 }
Otherwise, the fraction of semi-meanders with zero winding may be 
exponentially small,
and we only expect that ${R}(q) < {\bar R}(q)$ if $\nu(q)=1$.

\subsec{Exact results for large numbers of connected components
($q=\infty$)}

For very large $q$, we simply have 
\eqn\asyminf{ {\bar m}_n(q)~\sim ~ q^n }
as the meander polynomial is dominated by the $k=n$ term, 
corresponding to the unique semi-meander of order $n$ made of $n$ 
concentric circular roads, each crossing the semi-infinite river only once.
This semi-meander will appear as the rightmost object in the $n$-th
line of the tree of Fig.8.
The winding of this semi-meander is clearly $w=n$, hence we have, for
$q \to \infty$
\eqn\largq{ {\bar R}(q) \to q \qquad \gamma(q) \to 0 \qquad {\bar c}(q) \to 1 
\qquad \nu(q)\to 1}

As to meanders, the only way to build a meander of order $n$ 
with the maximal number $n$ connected components 
is that each component be a circle, 
crossing the river exactly twice. This is readily done by taking any upper
arch configuration and completing it by reflection symmetry w.r.t. the river.
This leads to $M_n^{(n)}=c_n$ (c.f. \catal) 
meanders with $n$ connected components. 
By Stirling's formula, we find that when $q \to \infty$ the meander polynomial 
behaves as
\eqn\asymqlm{ \eqalign{ m_n(q) ~ &\sim ~ c_n ~ q^n \cr
&\sim {1 \over \sqrt{\pi}} { (2 \sqrt{q})^{2n} \over n^{3/2} } \cr}}
hence, when $q \to \infty$
\eqn\asymqlmea{ {R}(q) \to 2 \sqrt{q} \qquad \alpha(q) \to 3/2 
\qquad {c}(q) \to 1 /\sqrt{\pi} }
This confirms the abovementioned property \equalir\ that ${R}(q)<{\bar R}(q)$
when $\nu(q)=1$, as $2\sqrt{q} < q$ for large $q$.

\subsec{Exact results for random walks on a half-line ($q=1$)} 

When $q=1$ in \semidef, ${\bar m}_n(1)$ simply counts all the multi-component
semi-meanders, irrespectively of their number of connected components. This
simplifies the problem drastically, as we are simply left with a purely
combinatorial problem which can be solved exactly.
The multicomponent semi-meanders are obtained by superimposing any arch 
configuration of order $n$ with the rainbow of order $n$, hence 
\eqn\qone{ {\bar m}_n(1)~=~c_n~\sim
{1 \over \sqrt{\pi}} {4^{n} \over n^{3/2}} }
by use of Stirling's formula for large $n$. 
This gives the values
\eqn\qoneval{ {\bar R}(1)=4 \qquad \gamma(1)=3/2 \qquad
{\bar c}(1)=1/\sqrt{\pi} }

The study of the winding at $q=1$ is more transparent in the
formulation of arch configurations of order $n$ as random walks of $2n$
steps on a semi-infinite line.  For each arch configuration of order
$n$, let us label by $1$, $2$,...,$2n-1$ each segment of river
in-between two consecutive bridges, and $0$ the leftmost semi-infinite
portion, $2n$ the rightmost one.  Let $h(i)$, $i=0,1,...,2n$ denote the
number of arches passing at the vertical of the corresponding segment
$i$. By definition, $h(0)=h(2n)=0$.  More generally, going along the
river from left to right, we have $h(i)=h(i-1)+1$ (resp.
$h(i)=h(i-1)-1$) if an arch originates from the bridge $i$ (resp.
terminates at the bridge $i$).

\fig{A walk diagram of $18$ steps, and the corresponding 
arch configuration of order $9$. Each dot corresponds to a segment of river. 
The height on the walk diagram is given by the number of arches 
intersected by the vertical dotted line.}{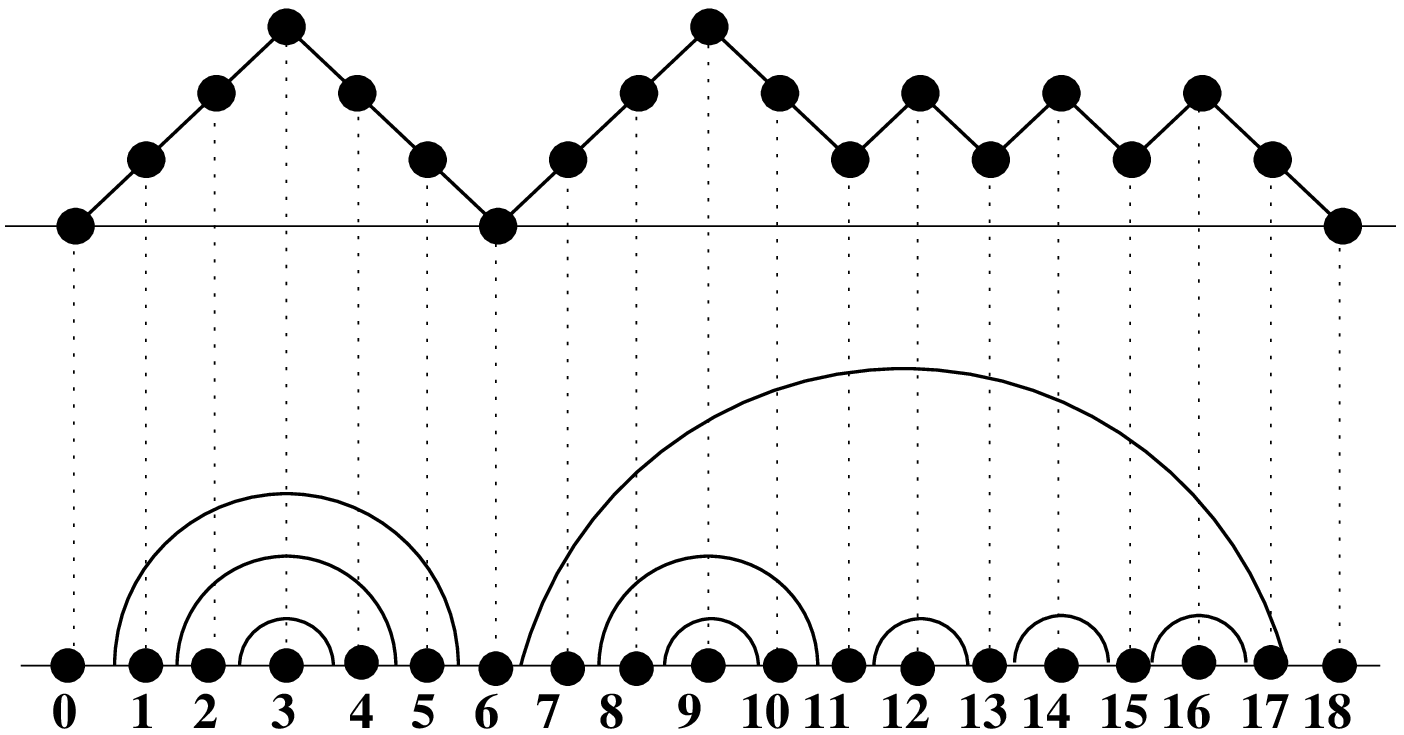}{9.cm}
\figlabel\walkyrie

The function $h$ satisfies $h(i)\geq 0$, for all $i$, and may be
interpreted as a ``height" variable, defined on the segments of river,
whose graph is nothing but a walk of $2n$ steps as shown in
Fig.\walkyrie.  This may be seen as the two-dimensional extent of a
brownian motion of $2n$ steps on a half-line, originating and
terminating at the origin of the line.  This interpretation makes the
leading behavior $c_n~\sim~2^{2n}$ of \qone\ clear:  it corresponds to
the $2$ possible directions (up or down) that the motion may take at
each step. The exponent $3/2$ in \qone\ is characteristic of the
boundary condition, namely that the motion is closed and takes place on
a half-line (other boundary conditions would lead to different values
of $\gamma$, e.g. for a closed walk on a line, we would have a behavior
${n \choose 2n}\sim 2^{2n}/\sqrt{n}$).

In this picture, the winding is simply given by the height $w=h(n)$ of 
the middle point.
Let us evaluate more generally the average height of a point $i$ over the
arch configurations of order $n$. It is given by 
\eqn\avhconf{\langle h(i) \rangle_n~=~{1 \over c_n}
\sum_{h \geq 0} h A_{n,i}(h)}
where $A_{n,i}(h)$ denotes the
number of arch configurations of order $n$ such that $h(i)=h$.
A simple calculation \NTLA\ shows that
\eqn\numofarc{ A_{n,i}(h)~=~
\left({i\choose {i+h \over 2}} -{i\choose {i+h \over 2}+1}\right)
\left({2n-i\choose n-{i-h \over 2}} -{2n-i\choose n-{i-h \over 2}+1}\right)}
as the $A_{n,i}(h)$ walks are simply obtained
by gluing two independent walks of $i$ and $2n-i$ steps linking the origin to 
the height $h$.

In the case of the winding, $w=h(i=n)$, \avhconf\
leads to a more compact formula, according to the parity of $n$
\eqn\resultqonewin{\eqalign{
n=2p:\ \ \ \ \ \ \  \langle w \rangle_{2p}~&
=~{{2p \choose p}^2 \over c_{2p}} -1\cr
n=2p+1:\ \ \ \ \langle w \rangle_{2p+1}~&=~
2{{2p \choose p}{2p+1 \choose p} \over c_{2p+1}} -1 \cr}}
For large $n$, this gives the following expansion
\eqn\stirone{  \langle w \rangle_{n}~=~2 \sqrt{n \over \pi}-1
+{5 \over 4 \sqrt{\pi n}}+ O(1/n^{3/2}) }
irrespectively of the parity of $n$. This implies that
\eqn\nuone{ \nu(q=1)~=~ 1/2 }
This is the well-known result for the Brownian motion, for
which the extent of the path scales like $n^{1/2}$ for large $n$. 
It is instructive to
note that, thanks to \stirone, the observable $w+1$ is less sensitive
than $w$ to the finite size effects at $q=1$. 
This will be useful in the forthcoming numerical estimates for 
arbitrary $q$ where we
observe that the numerical extrapolations are improved by considering $w+1$
instead of $w$.
Using \numofarc, we may now compute the probability distribution $P_n(w)$ 
for an arch 
configuration of order $n$ to have winding $h(n)=w$, which takes
for large $n$ the scaling form
\eqn\proba{ 
P_n(w)~=~{1 \over c_n} A_{n,n}(w)~
\sim~{1 \over \langle w \rangle_n}f\left({w \over \langle w \rangle_n}\right)}
with a scaling function $f$ independent of $n$ for large $n$,
readily obtained by use of Stirling's formula, upon writing
$w=2\sqrt{n/\pi}~\xi$ for large $n$. This gives 
\eqn\scafunc{ f(\xi)~=~ {32 \over \pi^2} \xi^2 e^{-{4 \over \pi}\xi^2}  }
for all $\xi>0$.

For general position $i\neq n$, we find, by a saddle point evaluation
of the sum \avhconf, that the average profile of
arch configurations is a ``Wigner" semi-circle
\eqn\semicirc{ \langle h(i) \rangle_n ~\sim~ 2 \sqrt {n\over \pi} \,
\sqrt{x(2-x)}}
when expressed in the scaled position $x=i/n$, $0 \leq x \leq 2$. 

\medskip

The meanders of order $n$
are the semi-meanders of order $2n$ with winding 
$w=h(2n)=0$. 
They are therefore built as the juxtaposition of two independent
walks of length $2n$. Hence
\eqn\meanone{ m_n(1)~=~ (c_n)^2 ~\sim ~ {1 \over \pi} {4^{2n} \over n^3} }
or, in other words
\eqn\resuone{ {R}(1)={\bar R}(1)=4 \qquad \alpha(1)=3 \qquad {c}(1)=1/\pi}
This is again in agreement with \equalir, as $\nu(1)=1/2<1$, i.e. the
winding is irrelevant at $q=1$.
 
\newsec{Exact enumeration and its analysis} 

In this section, we present results of an exact enumeration of ${\bar
M}_n^{(k)}$ for small $n$ ($n\leq 29$), and analyze their large $n$
extrapolation.  The enumeration is performed by implementing on a
computer a recursive algorithm which describes all the semi-meanders up
to some given order.  Clearly, the complexity is proportional to the
Catalan numbers ($c_n \sim 4^n$) hence the limitation on $n$.

\subsec{The main recursion relation}

The subsequent numerical study relies on the exploitation of the following
recursion relation generating all the semi-meanders of order $(n+1)$ from
those of order $n$.

\fig{The construction of all the semi--meanders of order $n+1$ 
with arbitrary number of connected components from those
of order $n$. Process (I): (i) pick any exterior arch and cut it (ii)
pull its edges around the semi--meander and paste them below.
The lower part
becomes the rainbow configuration ${\cal R}_{n+1}$ of order $n+1$.
This process preserves the number of connected components $k \to k$.
Process (II): draw a circle around the semi--meander of order
$n$. This process adds one connected 
component $k \to k+1$.}{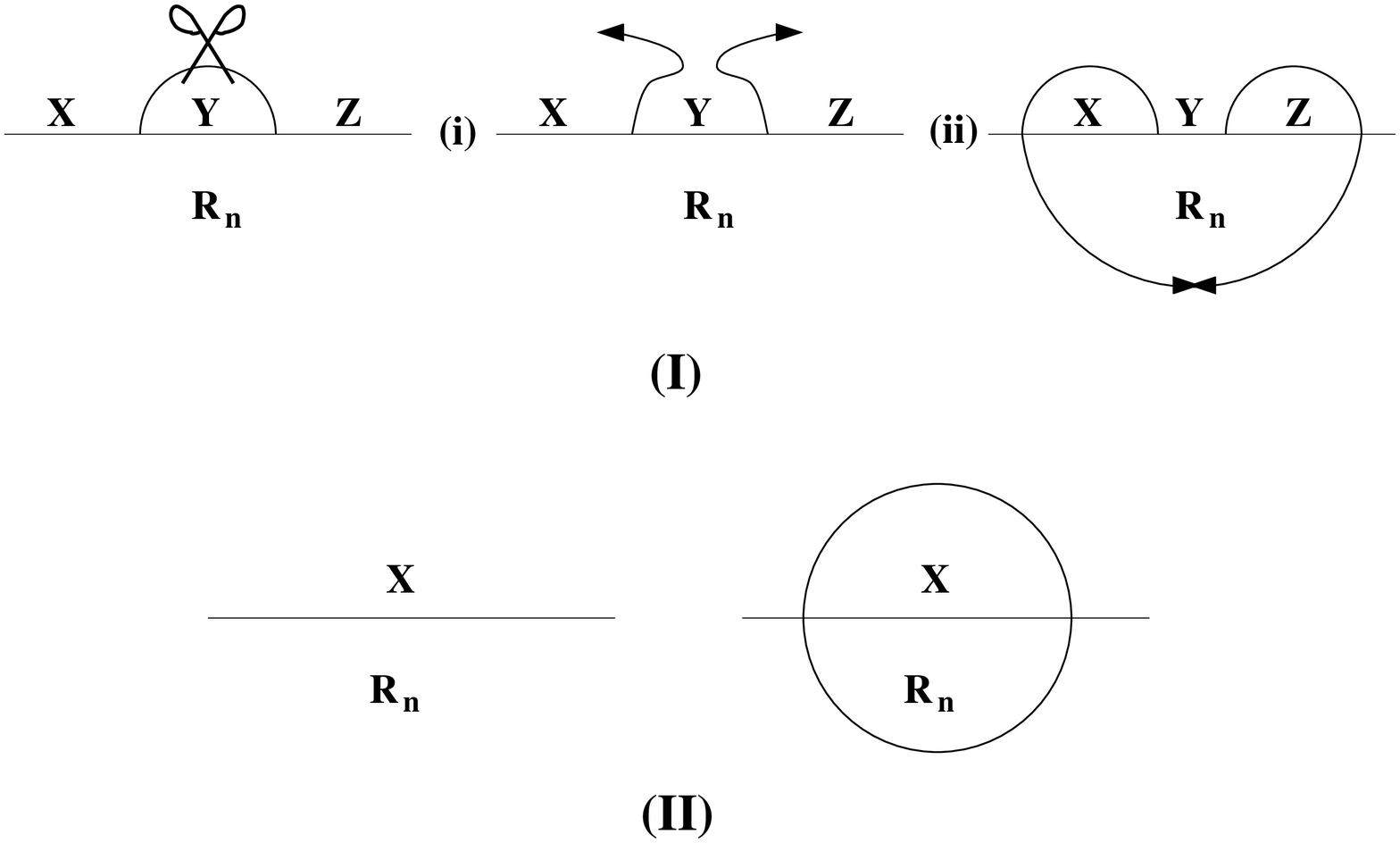}{11.truecm}
\figlabel\cutmeander

We start from any semi-meander of order $n$ with $k$ connected components,
in the open-river picture.
We may construct a semi-meander of order $(n+1)$ in either following way
(denoted (I) or (II)),
as illustrated in Fig.\cutmeander\

(I) Pick any exterior arch, i.e. any arch with no other arch passing above it.
Cut it and pull its ends all the way around the others (in order to add two 
bridges), and reconnect them below, by creating an
extra concentric lower arch for the rainbow.  
In this process, we have $n \to n+1$, but the number of connected components
has not changed: $k \to k$.  Another way of picturing this transformation
is the following: one simply has pulled the exterior arch all the way 
around the semi-meander and brought it below the figure, creating two new 
bridges along the way. As no cutting nor pasting is involved, the number 
of connected components is clearly preserved.

(II) Draw a circle around the semi-meander. This adds a lower concentric 
semi-circle which increases the order of the rainbow to $(n+1)$, and also
adds one connected component to the initial semi-meander $k \to k+1$.

These two possibilities exhaust all the 
semi-meanders of order $(n+1)$, as the transformation is clearly invertible,
by pulling back up the lower external arch of the rainbow.
Note that by construction, there are as many possibilities
for the process (I) as exterior arches, and 
the transformation is therefore one-to-many.

\fig{The tree of semi-meanders down to order $n=4$. 
This tree is constructed by repeated 
applications of the processes (I) and (II) on the semi-meander of order $1$
(root). We have indicated by small vertical arrows the multiple
choices for the process (I), each of which is indexed by its number.
The number of connected components of a given semi-meander is equal
to the number of processes (II) in the path going from the root 
to it, plus one (that
of the root).}{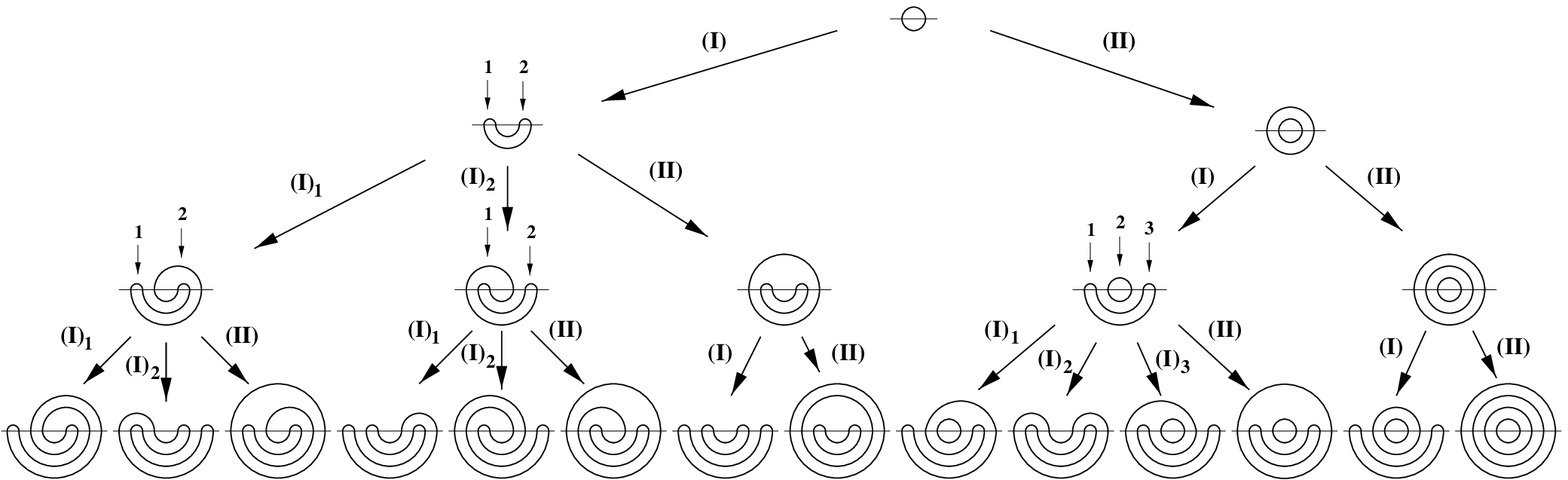}{12.cm}
\figlabel\treeofsm

\medskip

We may now construct a tree of all the semi-meanders, generated
recursively from that of order $1$ (root), as displayed in
Fig.\treeofsm.  Note that we have adopted the open-river formulation to
represent them.

Keeping track of the connected components, this translates into 
the following relation between the semi-meander polynomials
\eqn\recusemi{ {\bar m}_{n+1}(q)~=~ 
{\bar m}_n(q) \langle {\rm ext.arch.}\rangle_n(q) + q {\bar m}_n(q) }
where we denoted by $\langle {\rm ext.arch.}\rangle_n(q)$ the average
number of exterior arches in a semi-meander of order $n$, weighed by
$q^k$, $k$ its number of connected components.  In \recusemi, the first
term corresponds to all the processes (I), whereas the second term
corresponds to (II).

Taking the large $n$ limit in \recusemi, this permits to interpret
\eqn\extarch{{\bar R}(q)-q~=~\langle {\rm ext.arch.}\rangle_\infty(q)}
as the limit when $n \to \infty$ of the average number of exterior
arches in semi-meanders of order $n$, weighed by an activity $q$ per
connected component.

In particular, when $q \to \infty$, we have 
\eqn\limir{ {\bar R}(q)-q \to 1 } 
as the limiting
semi-meander, made of concentric circles, has only one exterior arch. 
This is a refinement of the large $q$ estimate in \largq.
When $q=1$, we find an average of
\eqn\avextone{\langle {\rm ext.arch.}\rangle_\infty(1)~=~
{\bar R}(1)-1~=~ 3 }
exterior arches in arbitrary arch configurations of large order \DGG.
When $q=0$, the connected semi-meanders, with $k=1$, are obtained 
through repeated action of the process (I) only.   
This restricts accordingly
the tree of Fig.\treeofsm. 
In that case, the partition function per bridge
\eqn\fperbri{ {\bar R}~=~ {\bar R}(0)~=~ \langle
{\rm ext.arch.}\rangle_\infty(0) }
coincides with the average number of exterior arches in connected
semi-meanders, for large $n$.
 
\subsec{Exact enumeration and large $n$ extrapolation}

In appendix A, we give an archetypical example of the programs 
we have implemented to compute the semi-meander numbers and various 
observables.
We have computed the numbers ${\bar M}_n^{(k)}$ for 
$1 \leq k \leq n \leq 27$, the numbers ${\bar M}_n^{(1)}$ for 
$n \leq 29$. To investigate the winding of semi-meanders,
we have also computed the numbers ${\bar M}_n^{(k)}(w)$ of semi-meanders
of order $n$, with $k$ connected components and fixed winding $w$ for
$1 \leq k \leq n \leq 24$ (encoding in particular the meander numbers
$M_n^{(k)}={\bar M}_{2n}^{(k)}(0)$), and the semi-meander ``profile" 
\eqn\prosm{ \sum_{{\rm semi-meanders} \atop
{\rm order }\ n, k\ {\rm c.c.}} h(i) }
for all positions $0 \leq i \leq 2n$.
Some of these numbers can be found in appendix A.

After gathering these numbers into generating functions of $q$, 
it is possible to
perform large $n$ extrapolations at fixed $q$, for the quantities 
${\bar R}(q)$, $R(q)$, $\gamma(q)$, $\alpha(q)$, ${\bar c}(q)$,
$c(q)$ and $\nu(q)$.

The general extrapolation scheme is the following. Suppose an observed
quantity $X_n$ has the following large $n$ expansion
\eqn\xnquant{ X_n~=~ \sum_{k=0}^p {x_k \over n^k} +O(1/n^{p+1}) }
Then we get a best estimate of the large $n$ limit $x_0$ by iterating
$p$ times the difference process $(\Delta f)(n)\equiv f(n+1)-f(n)$ 
on the function $f(n)=n^p X_n$, with the result
\eqn\resextrapo{{\Delta^p \over p!} \, n^p\, 
X_n ~=~ x_0 + O(1/n^{p+1})}
This gives perfect results for the Catalan numbers (i.e., $q=1$)
using $X_n={\rm Log}(c_{n+1}/c_n)$. This turns
out to extend to a whole range of $q$'s in a neighborhood of $1$.
For instance, ${\rm Log}\,{\bar R}(q)$ is extrapolated using 
$X_n={\rm Log}\sqrt{{\bar m}_{n+1}(q)/{\bar m}_{n-1}(q)}$.

\fig{The functions ${\bar R}(q)$ and $R(q)$ for $0 \leq q\leq 4$
as results of large $n$ extrapolations. The two curves coincide
for $0 \leq q\leq 2$ and split for $q>2$
with ${\bar R}(q)>R(q)$. Apart from the exact value ${\bar R}(1)=R(1)=4$, 
we find the estimates ${\bar R}(0)=3.50(1)$, ${\bar R}(2)=4.44(1)$,
${\bar R}(3)=4.93(1)$ and ${\bar R}(4)=5.65(1)$.}{rbb.eps}{8. cm}
\figlabel\rbarbare

The results for ${\bar R}(q)$ and $R(q)$ are displayed in Fig.\rbarbare.
The two functions are found to coincide in the range $0 \leq q \leq q_c$
with $q_c \simeq 2$, and to split into ${\bar R}(q)>R(q)$ for $q>q_c$.
As explained before, the comparison between ${\bar R}(q)$
and $R(q)$ determines directly whether $\nu(q)$ is $1$ or not. 
The result of Fig.\rbarbare\ is therefore the signal of a phase transition 
at $q=q_c$
between a low-$q$ regime where the winding is essentially irrelevant 
($\nu(q)<1$) and a large-$q$ phase with relevant winding ($\nu(q)=1$).

\fig{The winding exponent $\nu(q)$ for $0 \leq q \leq 8$,
as obtained from a large $n$ extrapolation. We observe a drastic change
of behavior between low $q$'s and large $q$'s, with an intermediate 
regime where the extrapolation fails, hence is not reliable. 
The dashed line indicates a possible scenario for the exact function
$\nu(q)$, compatible with a transition at $q_c\simeq 2$.
Apart from the exact value $\nu(1)=1/2$, we read 
$\nu(0)=0.52(1)$.}
{nuque.eps}{9. cm}
\figlabel\nuque

This is compatible with the direct extrapolation for $\nu(q)$ 
displayed in Fig.\nuque, which is however less reliable in the 
region around $q=2$, due to its sub-leading (and probably discontinuous)
character.

\fig{The configuration exponent $\gamma(q)$ for 
$0\leq q \leq 4$, from two different large $n$ extrapolations.
Apart from the exact value $\gamma(1)=3/2$, we estimate 
$\gamma(0)\simeq 2$.}{gamma.eps}{8.cm}
\figlabel\gaga

The configuration exponent for 
semi-meanders $\gamma(q)$ is represented in Fig.\gaga,
for two different orders in the extrapolation scheme \resextrapo.
The extrapolation proves to be stable for $0<q <2$. For
$q>2$, it develops oscillations around a mean value, estimated to 
vanish ($\gamma(q) \sim 0$) for $q$ large enough. 
For simplicity,
we chose not to represent the functions $\alpha(q)$, $c(q)$, ${\bar c}(q)$.
The coefficient 
${\bar c}(q)$ develops a discontinuity at the transition
$q=2$.
On the other hand,  
the functions pertaining to meanders only ($\alpha(q)$ and $c(q)$)
do not display any transition at $q=2$.

\subsec{Scaling functions}

By analogy with critical phenomena, in addition to 
the scaling behaviors \asympto, \asymean\ and \defwik\ involving
the critical exponents $\gamma(q)$, $\alpha(q)$ and $\nu(q)$,
we expect to find more refined scaling laws involving
scaling functions.  
A particular example of such scaling functions has been
derived for $q=1$ \proba, for the probability distribution
$P_n(w)$ of the winding $w$ among arch configurations of order $n$.
It involves the scaling function \scafunc.
For $q=0$ we expect the same behavior for the
corresponding probability distribution 
\eqn\defprom{P_n^{(0)}(w)~=~ 
{{\bar M}_n^{(1)}(w) \over {\bar M}_n^{(1)}} }
of winding $w$ among 
connected semi-meanders of order $n$.
We expect the scaling behavior
\eqn\scacom{ P_n^{(0)}(w)~ \sim~ {1 \over \langle w \rangle_n(0)}
f^{(0)}\left( {w \over  \langle w \rangle_n(0)}\right) }

\fig{Plot of $\langle w+1 \rangle_n(0)\,  P_n^{(0)}(w)$ as 
a function of the reduced variable 
$\xi=(w+1)/\langle w +1\rangle_n(0)$ for $n=2,3,...,24$.
The points accumulate to a smooth scaling function $f^{(0)}(\xi)$.
The erratic points correspond to small values of $n$, 
which have not reached the
asymptotic regime.}{histo0.eps}{6.5cm}
\figlabel\histzero

This is precisely what we observe in Fig.\histzero, where we plot
$\langle w+1 \rangle_n(0)\,  P_n^{(0)}(w)$ as a function of the reduced 
variable $\xi=(w+1)/\langle w +1\rangle_n(0)$ for different values of $n$.
Indeed, as already explained in the $q=1$ case, we have taken the 
variable $(w+1)$ instead of $w$ to improve the convergence.
All the data accumulate on a smooth curve, which represents
the scaling function $f^{(0)}(\xi)$. The shape of this function is 
reminiscent of that of the end-to-end distribution for polymers.
By analogy, we expect a certain power law behavior for small $\xi$
\eqn\smabeh{ f^{(0)}(\xi) ~\sim~ \xi^\theta }
where $\theta$ satisfies the relation
\eqn\relathet{ \alpha -\gamma~=~ \nu (1+\theta) }
obtained by identifying
\eqn\proport{ P_{2n}^{(0)}(0)~\propto~ {1\over n^\nu} f^{(0)}\left({1
\over n^\nu}\right)}
to 
\eqn\orprop{ {{\bar M}_{2n}^{(1)}(0)\over {\bar M}_{2n}^{(1)}}~=~
{M_{n}\over {\bar M}_{2n}} \propto n^{\gamma-\alpha}}
For large $\xi$, we expect a behavior
$f^{(0)}(\xi) \sim \exp(-{\rm const.}\, \xi^\delta)$ with a possible
Fisher-law behavior $\delta=1/(1-\nu)$. The observed function
of Fig.\histzero\ is compatible with these limiting behaviors, 
although we cannot extract reliable estimates of the exponents
$\theta$ and $\delta$.

As we already did in the case of $q=1$ \semicirc, we can study 
the average profile of semi-meanders 
\eqn\avprofsm{ {\langle h(i) \rangle_n(q) \over
\sum_{j=0}^{2n}\langle h(j) \rangle_n(q) }~\sim~ 
{1 \over n} \rho(x={i/n};q) }
involving a scaling function $\rho(x;q)$ of the variable $x$,
with $0 \leq x \leq 2$ for each value of $q$
(with the appropriate normalization
such that $\int \rho =1$).
For instance, we have seen in eq.\semicirc\ that
$\rho(x;1)= (2/\pi) \sqrt{x(2-x)}$.

\fig{Semi-meander average profiles for $q=0$, $1$, $2$, $4$,
and $1 \leq n \leq 24$, as functions of the reduced 
variable $x=i/n$. For $q=1$, we also represented the exact 
large $n$ Wigner semi-circular limit $\rho(x;1)$. 
For $q=4$, we also
represented the large $n$ and $q$ piecewise-linear 
limit $\rho(x;\infty)$.}{profils.eps}{9.cm}
\figlabel\profils

We have represented in Fig.\profils\ these profiles for several 
values of $q$. Again, the points accumulate on smooth
limiting curves $\rho(x;q)$. We observe a first change of behavior at $q=1$
between a $q<1$ regime with a negative cusp at $x=1$ and a 
$q>1$ regime with a positive cusp, separated by the Wigner
semi-circle, with no cusp at $q=1$. For large $q$, $\rho(x;q)$ tends
to the limit $\rho(x;\infty)=1-|1-x|$ corresponding to the unique semi-meander
made of $n$ concentric circles, which satisfies $h(i)=i$ for 
$0 \leq i\leq n$ and $h(i)=2n-i$ for $n \leq i \leq 2n$.
For small $x$, we expect a power law behavior
of the form 
\eqn\polaw{ \rho(x;q)~\sim ~x^{\varphi(q)} }
where we identify the exponent $\varphi(q)=\nu(q)$ from $h(1)=1$ and
the fact that $\sum_j h(j) \sim n^{1+\nu(q)}$.

\newsec{Large $q$ asymptotic expansions}

In the previous section, we have observed two regimes
for the semi-meander polynomials, namely a low-$q$ regime in
which the winding is irrelevant and a large-$q$ regime where
the winding is relevant, separated by a transition at a value of
$q=q_c\simeq 2$.
On the other hand, we have already exhibited an exact solution
of the problem at $q=\infty$ \largq, and a first correction thereof 
for large $q$ in \limir. It is therefore tempting to analyze the
large $q$ phase by a systematic expansion
in $1/q$.
This is performed in the following section, where ${\bar R}(q)$
is expanded up to order $19$ in $1/q$, and $\gamma(q)$ is found
to vanish identically throughout the
large $q$ regime.
In the subsequent section, we compute the large $q$ expansion
of the average winding in semi-meanders, and
we find $\nu(q)=1$ identically in this regime.
 
\subsec{Large $q$ asymptotic expansion of the semi-meander polynomial}

In this section we derive the large $q$ expansion of the semi-meander
polynomial ${\bar m}_n(q)$ of eq.\semidef\ as
\eqn\asypolq{{\bar m}_n(q)~=~ q^n~({\bar M}_n^{(n)} + 
{{\bar M}_n^{(n-1)}\over q}
+{{\bar M}_n^{(n-2)}\over q^2} + \cdots ) }
involving the semi-meander numbers in the
form ${\bar M}_n^{(n-k)}$, $k=0,1,2$,... 
Remarkably, these numbers, for arbitrary $n\geq 2k-1$,
are polynomials of $n$, which furthermore exhibit 
some special structure allowing for
an explicit large $q$ expansion of ${\bar R}(q)$.

The section is organized as follows. We first derive 
the polynomial form of the ${\bar M}_n^{(n-k)}$, valid
for $n\geq 2k-1$, together with the corrections to be added 
for smaller $n$'s.

The re-exponentiation of ${\bar m}_n(q)$ in the form \asympto\ 
induces strong constraints on the polynomials ${\bar M}_n^{(n-k)}$,
which allow for their complete determination up to $k=18$,
out of their first values for small $n$, which were enumerated
exactly up to $n=27$.

\medskip
 
For starters, let us first compute the numbers ${\bar M}_n^{(n-k)}$ 
for $k=0,1,2$.

\fig{Semi-meanders with many connected components. $(a)$ $k=n$ connected 
components;
there are $n$ circles. $(b)$ $k=n-1$ connected components; 
there are $(n-2)$ circles and one ``kidney". 
$(c)$ $k=n-2$ connected components; there are respectively
$(c)_1$ two disjoint kidneys and $(n-4)$ circles; $(c)_2$ two kidneys included
in one another and $(n-4)$ circles; $(c)_3$ and $(c)_4$ one ``spiral"
and $(n-3)$ circles.}{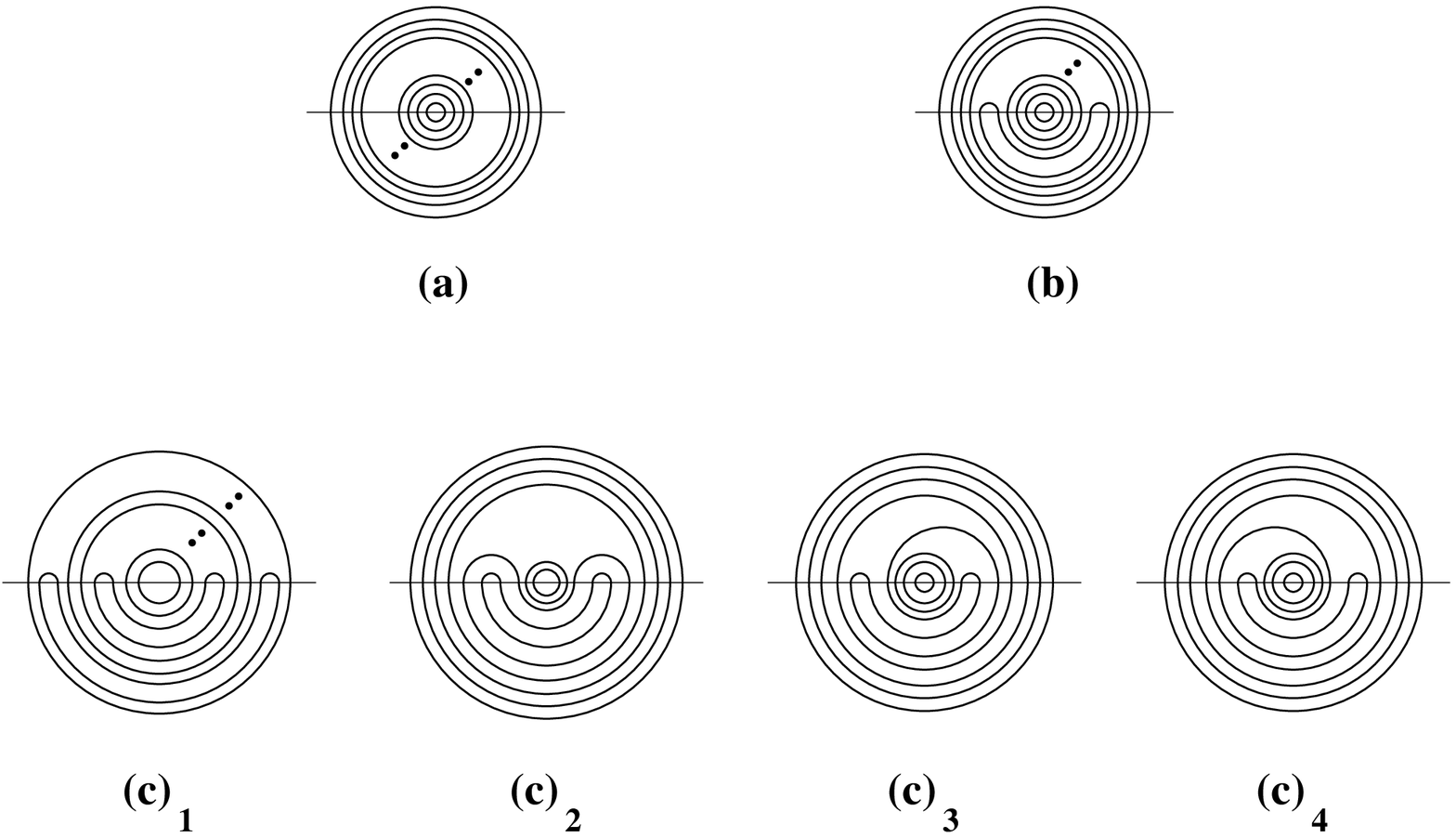}{10.cm}
\figlabel\firstorder

As we already mentioned, the leading term ${\bar M}_n^{(n)}=1$
in the expansion \asypolq\ counts the 
unique semi-meander of order $n$ with $n$ connected components 
made of concentric circles only, and which we refer to as 
the leading semi-meander
(c.f. Fig.\firstorder\ (a)).  
This yields the first polynomial 
\eqn\pofi{p_0(n)~=~{\bar M}_n^{(n)}~=~1}
for all $n \geq 0$.

The sub-leading term is made of the 
\eqn\kidneyper{p_1(n)~=~{\bar M}_n^{(n-1)}~=~n-1} 
``kidney''-type perturbations of the leading
semi-meander, displayed in Fig.\firstorder\ (b).

The next-to-leading term consists of the semi-meanders of order $n$
with $(n-2)$ connected components, which may be obtained as follows. 

(i) A first possibility
consists in taking two ``kidney''-type perturbations of the leading
semi-meander (see Figs.\firstorder\ $(c)_1$ and $(c)_2$), which are either

\item{}(i1) (Fig.\firstorder\ $(c)_1$) disjoint, hence a total of 
$(n-2)(n-3)/2$ choices for the positions of two kidneys.

\item{}(i2) (Fig.\firstorder\ $(c)_2$) included in one another, 
hence a total of $(n-3)$ choices
for the position of the double kidney, or $0$ choice if $n=2$.

(ii) The second possibility is a larger ``spiral''-type perturbation of the
leading semi-meander, with a total of $(n-2)$ available positions, and
there are two such perturbations (see Figs.\firstorder\ $(c)_3$
and $(c)_4$).

Summing up all these contributions gives\foot{Throughout this 
section, we use the fact that ${\bar M}_n^{(0)}=\delta_{n,0}$,
i.e. there are no semi-meanders with zero connected component, 
except for the vacuous semi-meander of order $0$.} 
\eqn\pertum{ {\bar M}_n^{(n-2)}~=~ {(n-2)(n-3)\over 2} +(n-3)+\delta_{n,2}
+2(n-2)~=~ {n^2+n-8 \over 2} + \delta_{n,2} }
hence a polynomial 
\eqn\posec{ p_2(n)~=~{n^2+n-8 \over 2} }
We see here the first appearance of a correction for small $n$,
in that an extra boundary term ($\delta_{n,2}$) has to be added at $n=2$, 
to recover the fact that ${\bar M}_2^{(0)}={0}$.

More generally, let us consider the number ${\bar M}_n^{(n-k)}$
for large $n$ and finite $k$. The corresponding semi-meanders are obtained
in the tree of Fig.\treeofsm\ by applying to the root $k$ times 
the process (I)
and $(n-1-k)$ times the process (II). 
This gives ${n-1 \choose k}\sim n^k/k!$
possible choices ($k<<n$). These choices however are not 
completely independent. 

Recall that the process (I) may be applied to any exterior arch of the 
semi-meander. In the situation where $k<<n$, the semi-meander will most
probably have only one exterior arch (generated by the last process,
most probably of the type (II)), and there will be only 
one choice for (I), creating a kidney. 
If two or more processes (I) are applied 
{\it consecutively}, the number of exterior arches may grow, 
and yield more semi-meanders (e.g. the application of two
consecutive processes (I) yields three possibilities:
two included kidneys, or any of the two spirals). 
Such an effect is however sub-leading, as the number of choices
of two or more consecutive processes (I) grows at most like $n^{k-1}$.
Collecting all these extra combinatorial factors permits, like in \pertum,
to write the number ${\bar M}_n^{(n-k)}=p_k(n)$ as a polynomial of 
degree $k$ in $n$
for large enough $n$, and fixed $k$, with moreover a leading term
$n^k/k!$.

For smaller $n$,
the above consecutive choices are affected by the boundaries of the tree.
Recall for instance that the double included kidneys of
Fig.\firstorder\ $(c)_2$, obtained by two successive applications of
(I) on two concentric circles, may only exist for $n \geq 4$. We have
therefore needed to specify that their number (generically equal to
$(n-3)$) vanishes for $n=3$ (granted) and $n=2$,
the latter resulting in a boundary correction $\delta_{n,2}$.
More generally, expressing that the combinatorial expressions 
found are only valid for large enough $n$'s will
translate into boundary terms, studied in detail in appendix B.
We simply quote the result here, valid for all $n \geq k$
\eqn\asypolsm{ {\bar M}_n^{(n-k)}~=~ p_k(n) + 
\sum_{j=0}^{k-2} {\bar \mu}_j^{(k)}\, \delta_{n,2k-2-j} }
where the ${\bar \mu}_j^{(k)}$ are some positive integers. 
In particular, \asypolsm\ shows that the formula ${\bar M}_n^{(n-k)}$
is a pure polynomial $p_k(n)$ with no corrections
as soon as $n \geq 2k-1$. This property is derived in appendix B,
where the first correction $\mu_0^{(k)}=c_{k-1}$ is also obtained. 

The precise determination of the $p_k$'s and the $\mu_j^{(k)}$ could
be in principle achieved directly by pursuing the above method
used for $k=0,1,2$. However the complexity of this program is
comparable to that of the exact enumeration of the semi-meander numbers.
Instead we can guess the coefficients
of $p_k(n)$ by matching our data for ${\bar M}_n^{(n-k)}$ with
the form \asypolsm. This can be pushed further by exploiting the
re-exponentiation property of ${\bar m}_n(q)$, which implies
relations between the coefficients of the $p_k(n)$, as discussed now. 

The property \asypolsm\ must be reconciled with the large $n$ behavior
of ${\bar m}_n(q)$ \asympto, namely that for $n$ sufficiently large
\eqn\propmnk{ {\rm Log}\, {\bar m}_n(q)~=~ n \,{\rm Log}\, {\bar R}(q)-
\gamma(q) {\rm Log}\, n + {\rm Log}\, {\bar c}(q) + o(1)}
Such an expansion is valid for all $q$, and we can in particular 
study it for large $q$.
On the other hand, up to any order $k$ in the $1/q$ expansion,
and by choosing $n\geq 2k-1$, 
we may also write
\eqn\pmnk{\eqalign{ 
{\rm Log}\, {{\bar m}_n(q)\over q^n}~&=~
{\rm Log} \big(1 + {p_1(n) \over q} + {p_2(n) \over q^2}+ 
\cdots + {p_k(n) \over q^k}+O( {1 \over q^{k+1} } ) \big) \cr
&=~\sum_{m=1}^k {1 \over q^m} \sum_{j=1}^m {(-1)^{j-1} \over j} 
\sum_{k_1+...+k_j=m\atop k_i \geq 1} 
p_{k_1}(n) p_{k_2}(n)...p_{k_j}(n) +O({1 \over q^{k+1}}) \cr}}
In this expansion, the coefficient of $1/q^m$
is a polynomial of $n$, as a sum of products of polynomials of $n$.
Comparing this with an expansion of \propmnk\ in $1/q$, 
we see that its degree is at most $1$.
Therefore, there exist two sequences of coefficients 
$(\alpha_k,\beta_k)$, such that
\eqn\constrpol{ p_k(n)~=~ \alpha_k n +\beta_k +
\sum_{j=2}^k {(-1)^j \over j} \sum_{k_1+...+k_j=k\atop
k_i \geq 1} p_{k_1}(n) p_{k_2}(n)...p_{k_j}(n) }
with the correspondence
\eqn\corpol{\eqalign{
{\rm Log}\, {\bar R}(q)~&=~ {\rm Log}\,q+\sum_{k\geq 1} {\alpha_k \over
q^k} \cr
{\rm Log}\, {\bar c}(q)~&=~ \sum_{k \geq 1} {\beta_k \over q^k} \cr}}
Moreover,
there can be no ${\rm Log}\, n$ term in the expansion of \propmnk, hence
the remarkable result 
\eqn\conseqga{\gamma(q)~=~0}
This result is expected to hold as long as the corrections 
to the polynomial behavior of the ${\bar M}_n^{(n-k)}$ are negligible.
As we will see, this condition defines precisely the
large $q$ phase $q>q_c$. Therefore the exponent 
$\gamma(q)$ vanishes identically over the whole phase $q>q_c$.
In view of, say, the exact value $\gamma(q=1)=3/2$,
this property cannot persist in the small $q<q_c$ phase.
This is not surprising since the first correction $\mu_0^{(k)}=c_{k-1}$
(for $k=n/2+1$)
implies an additional power law correction of the form $1/n^{3/2}$.

For $n\geq 2k-1$, \constrpol\ is a quasi-recursion relation for
the polynomials $p_k$, hence for the semi-meander numbers ${\bar M}_n^{(n-k)}$.
This relation is exploited in appendix C to generate from our numerical data
the polynomials $p_k(x)$ for $0 \leq k \leq 18$, together with 
their corrections.
Using these polynomials, it is now straightforward to
read the functions ${\bar R}(q)$
and ${\bar c}(q)$ using \corpol, with the result
\eqn\resultasympto{\encadremath{\eqalign{
{\bar R}(q)~&=~ q+1+{2 \over q}+{2 \over q^2}
+{2 \over q^3}-{4 \over q^5}
-{8 \over q^6}-{12 \over q^7}-{10 \over q^8}-{4 \over q^9}
+{12 \over q^{10}}
+{46 \over q^{11}}\cr
&+{98 \over q^{12}}+{154 \over q^{13}}+{124 \over q^{14}}+{10\over q^{15}}
-{102 \over q^{16}}+{20 \over q^{17}}-{64 \over q^{18}}
+O({1 \over q^{19}})\cr
{\bar c}(q)~&=~1-{1 \over q}-{4 \over q^2}
-{4 \over q^3}+{14 \over q^5}
+{44 \over q^6}+{56 \over q^7}+{28 \over q^8}-{82 \over q^9}
-{252 \over q^{10}}-{388 \over q^{11}}\cr
&-{588 \over q^{12}}-{772 \over q^{13}}-{620 \over q^{14}}
+{1494 \over q^{15}}+{5788 \over q^{16}}+{7580 \over q^{17}}
-{690 \over q^{18}}+O({1 \over q^{19}})\cr}}}

It is interesting to compare the result of these large $q$ expansions
to the previous direct large $n$ extrapolations of sect.3. As far
as
${\bar R}(q)$ is concerned, we find a perfect agreement for the values
$q \geq 2$, down to $q=2$, where we find ${\bar R}(2)\simeq 4.442(1)$
using \resultasympto,
in perfect agreement with the previous estimate. 
The precision of \resultasympto\
increases with $q$, leading to far better estimates than before:
${\bar R}(3)\simeq 4.92908(1)$, ${\bar R}(4)\simeq 5.6495213(1)$...

As to $\gamma(q)$, our prediction that $\gamma(q)=0$ for all $q>2$ is
compatible
with the previous extrapolation of Fig.\gaga, where this value
is represented in dashed line.  We therefore expect $\gamma(q)$
to have a discontinuity at $q=2$, where it goes from a non-zero 
$\gamma(q=2^-)$ value to zero.  This will be confirmed by the 
forthcoming analysis of the low-$q$ phase in sect.5.

\subsec{Large $q$ asymptotic expansion of the semi-meander winding}

In this section, we first examine the contribution of the circles to the 
winding of semi-meanders in the large $q$ phase.  It turns out to be
of the order $n$ throughout this phase, implying that
the winding exponent $\nu(q)=1$ for all $q>q_c$.
In a second step, we compute the large $q$ expansion of the
average winding by techniques similar to those of the previous section,
showing that the circles contribute only for a finite fraction of the 
winding in the large $q$ phase.
This study will single out the value $q_c=2$ with a very good precision.

To enumerate the total number of circles in 
order $n$ semi-meanders, we simply have to
count the semi-meanders with a marked circle, in one-to-one correspondence
with pairs of semi-meanders of total order $(n-1)$ (since the marked
circle separates the original meander into two disconnected pieces, its
inside and outside). 
Hence, the average number of circles
in semi-meanders is given by
\eqn\avcircsm{ \langle {\rm circ.} \rangle_n(q)~=~ q 
{\sum_{j=0}^{n-1} {\bar m}_j(q)\, {\bar m}_{n-1-j}(q)  \over {\bar m}_n(q)} }
valid for all $q$ and $n$, with the convention that ${\bar m}_0(q)=1$.
Using the large $n$
asymptotics ${\bar m}_n(q)\sim {\bar c}(q) {\bar R}(q)^n$, with $\gamma(q)=0$
throughout the large $q$ regime, we find
\eqn\asymcirc{  \langle {\rm circ.} \rangle_n(q)~\sim n\, 
q {{\bar c}(q) \over {\bar R}(q)}}
The average number of circles therefore grows like $n$ which in turn
implies that
\eqn\nuegone{ \nu(q)~=~1 \qquad q~>~q_c}
throughout the large $q$ regime,
since each circle contributes $1$ to the winding and clearly 
$\langle w\rangle \geq \langle {\rm circ.} \rangle$.
Note that the above argument relies crucially on the
fact that $\gamma(q)=0$, and thus cannot
be applied to the small $q$ regime. Indeed, for $q=1$, \avcircsm\ simply gives
$\langle {\rm circ.} \rangle_n(q=1)=1$ for all $n$, hence a very different
behavior.

The quantity $\displaystyle\sigma(q)=\lim_{n\to 
\infty}\langle {\rm circ.} \rangle_n(q)/n$ 
is simply given by the
Taylor expansion of $q {\bar c}(q)/{\bar R}(q)$
\eqn\circasympto{\eqalign{
\sigma(q)~&=~ 1-{2 \over q}-{4 \over q^2}+{2 \over q^3}+{8 \over q^4}
+{14\over q^5}+{22\over q^6}-{14\over q^7}-{66\over q^8}-{98\over q^9}
-{54\over q^{10}}\cr
&+{106\over q^{11}}+{20 \over q^{12}}-{282 \over q^{13}}-{220\over q^{14}}
+{1602\over q^{15}}+{3428\over q^{16}}
-{1330\over q^{17}}-{13824\over q^{18}}+O({1\over q^{19}})\cr}}

\medskip

Let us now turn to the large $q$ expansion of the
average winding of semi-meanders \avwind. 
This requires a refined
study of the semi-meander numbers ${\bar M}_n^{(n-k)}(w)$
with fixed winding $w$, which display a similar polynomial
structure as the ${\bar M}_n^{(n-k)}$. 
The study of the corresponding
generating function is presented in appendix D and leads to
\eqn\asymwin{ \langle w \rangle_n(q)~=~ \lambda(q) n +\mu(q) }
where the coefficients $\lambda(q)$ and $\mu(q)$ have the following
large $q$ expansions up to order $14$ in $1/q$
\eqn\lamu{\encadremath{\eqalign{ 
\lambda(q)~&=~ 1 -{2 \over q} -{2 \over q^2}+{2 \over q^3} 
+{2 \over q^4}+ {2 \over q^5}+{10 \over q^6}
-{6 \over q^7}-{14 \over q^8}-{10 \over q^9}\cr
&+{22 \over q^{10}}+{86 \over q^{11}}-{58 \over q^{12}}
-{222\over q^{13}}-{118\over q^{14}}+O({1 \over q^{15}})\cr
\mu(q)~&=~ {2 \over q} +{10 \over q^2}+ {22 \over q^3} 
+{54\over q^4}+{134\over q^5}+{246\over q^6}
+{622\over q^7}+{1434\over q^8}+{3178\over q^9}\cr
&+{6834\over q^{10}}+{13786\over q^{11}}+{30834\over q^{12}}
+{66590\over q^{13}}+{140582\over q^{14}}
+O({1 \over q^{15}})\cr}}}

\fig{The series $\lambda(q)$ \lamu\ and $\sigma(q)$ 
\circasympto\ of $1/q$
up to order $14$ and $18$ respectively, for $1<q<8$.
Both curves seem to vanish at $q=2$.}{lasig.eps}{8cm}
\figlabel\plolamsig

It can be checked directly that $\lambda(q)>\sigma(q)$ hence the
circles only contribute for a finite fraction of the total winding. 
The plots of the functions $\lambda(q)$ and $\sigma(q)$ 
are displayed in Fig.\plolamsig. 
Remarkably, both coefficients seem to vanish at the same point $q=2$
with an excellent precision.
Since these coefficients must be positive, we deduce that our
large $q$ formulas break down for $q<2$.
We interpret this as yet another evidence of the drastic 
change of behavior of 
the average winding $\langle w \rangle$, which is no longer 
linear in $n$ below $q_c$, and we find $q_c=2$ with an 
excellent precision. 

The actual break-down of the large $q$ phase is studied more
systematically in next section.

\newsec{The break-down of the large $q$ expansion for $q<q_c$}

The properties of the large $q$ phase are intimately based on the 
polynomial structure of the numbers ${\bar M}_n^{(n-k)}$. 
In this section, we explain the break-down of this phase
by the precise structure of the non-polynomial corrections 
\asypolsm\ to this behavior. The phase transition occurs 
when these corrections become dominant.  The detailed analysis of
these corrections shows a strong resemblance between the low $q$ 
phase and a meander (zero-winding) regime.
 
The corrections to the polynomial behavior of the ${\bar M}_n^{(n-k)}$
are gathered in the function
\eqn\defmupl{{\bar \mu}_n(q)~=~ \sum_{k=0}^{n}
\big({\bar M}_n^{(n-k)} - p_k(n) \big) q^{n-k}~
=~ \sum_{k=[(n+2)/2]}^{n}  {\bar \mu}_{2k-2-n}^{(k)} \, q^{n-k}}
The numbers ${\bar \mu}_{2k-2-n}^{(k)}$ are listed 
for $n/2 < k <n =2,3,...,9$ in appendix C. 
With the only difference that we are now dealing with rational fractions
of $n$ instead of polynomials, we can carry the same large $n$ expansion
as in sect.4.1, to extract the 
large $n$ asymptotics of ${\bar \mu}_n(q)$, with the result,
according to the parity of $n$
\eqn\respamu{\eqalign{
{\bar \mu}_{2n}(q)~&\sim~ {\bar c}_+(q)\, 
{{\bar R}_1(q)^{2n} \over (2n)^{\gamma_1(q)}} \cr
{\bar \mu}_{2n-1}(q)~&\sim~ {\bar c}_-(q)\, 
{{\bar R}_1(q)^{2n-1} \over (2n-1)^{\gamma_1(q)}} \cr}}
Using the results of appendix C, we get the following large $q$
expansions (incorporating the large
$n$ asymptotics of $c_n \sim 4^n/(\sqrt{\pi} n^{3/2})$)
\eqn\largqmsd{\encadremath{\eqalign{
{\bar R}_1(q)~&=~2 \sqrt{q} \big(1+{1 \over q}+{3 \over 2 q^2}-{3 \over 2 q^3}
-{29 \over 8 q^4} +O({1  \over q^5})\big) \cr
{\bar c}_+(q)~&=~ {2 \over q} \sqrt{2 \over \pi} 
\big( 1+{23 \over q}+{283 \over q^2}+{3027 \over q^3}+{313751 \over q^4} 
+O({1 \over q^5}) \big)\cr
{\bar c}_-(q)~&=~ {2 \over q} \sqrt{2 \over \pi q}
\big( 3+{40 \over q}+{417 \over q^2}+{4418 \over q^3}+{44991 \over q^4}
+O({1 \over q^5}) \big)\cr}}}
Due to the rational fractions of $n$ we have dealt with,
we cannot conclude that $\gamma_1(q)=3/2$ (the $3/2$ comes from the
Catalan number asymptotics) identically for large $q$. We expect this
to hold only in the $q \to \infty$ limit, whereas for finite $q$,
$\gamma_1(q)$ is some function of $q$.

We can now write 
\eqn\posmic{{\bar m}_n(q)~\simeq~{\bar c}_0(q)\, {\bar R}_0(q)^n 
+{\bar c}_{\pm}(q) {{\bar R}_1(q)^{n} \over n^{\gamma_1(q)} } + \cdots }
where $\pm$ is chosen according to the parity of $n$ and
we have indexed by $0$ the functions ${\bar R}(q)$ and ${\bar c}(q)$
corresponding to the polynomial contributions to the ${\bar M}_n^{(n-k)}$,
with the asymptotic expansions \resultasympto.
When $q$ is large, the first term dominates as ${\bar R}_0(q)\sim q$,
whereas ${\bar R}_1(q)\sim 2\sqrt{q}$.
This justifies a posteriori the identification ${\bar R}(q)={\bar R}_0(q)$
in the large $q$ regime of sect.4.
The properties derived in sect.4. for the large $q$ regime
will however break down if ${\bar R}_1(q)\geq {\bar R}_0(q)$, i.e. when
the corrections become dominant.
We expect such a crossing of phases to take place at the transition
point $q=q_c$, at which ${\bar R}_1(q_c)={\bar R}_0(q_c)$.
In this scenario, ${\bar R}(q)={\bar R}_1(q)$ for all $q<q_c$,
and the exponent $\gamma(q)$ jumps abruptly from its large $q>q_c$
value $\gamma_0(q)=0$, to a non-zero value $\gamma_1(q_c)\neq 0$. 
This mechanism explains the jump in the value of 
$\gamma(q)$ at the vicinity of $q=q_c$, as observed in Fig.\gaga.
In this figure, we have represented in dashed line the purported 
value $\gamma(q)=0$ for $q>2$, which must be substituted to the
(bad) large $n$ estimate in this regime, whereas we still
rely on the (good) $q<2$ estimate.

To reconcile this scenario with the picture described
in sect.3, in which for $q<q_c$ 
the winding of semi-meanders becomes irrelevant (i.e. ${\bar R}(q)=R(q)$),
we should have
${\bar R}_1(q)=R(q)$ below the transition point $q_c$, where
the semi-meanders enter their meander-like phase.  
Since the meanders themselves do not display any transition at $q=q_c$,
it is tempting to infer that ${\bar R}_1(q)=R(q)$ for all values $q$.
As we will see now, this is corroborated by the large $q$
expansion \largqmsd\ for ${\bar R}_1(q)$, which turns out to coincide 
with that of $R(q)$ for meanders \asymean.
The latter is easily carried out 
for the meander numbers,
whose structure, given in appendix E, is very similar to that 
of the semi-meander corrections. 
More precisely, the meander numbers $M_n^{(n-k)}$ take the general form
\eqn\geneform{ M_n^{(n-k)}~=~ c_n \, r_k(n) }
where $r_k(x)$ is a rational fraction of $x$, with total degree $k$.
With the values listed in appendix E, we find the following 
large $q$ expansion for the functions $R(q)$ and $c(q)$ of \asymean,
up to order $6$ in $1/q$
\eqn\largqmean{\encadremath{\eqalign{
R(q)~&=~2 \sqrt{q} \big(1+{1\over q}+{3\over 2 q^2}-{3\over 2 q^3}
-{29\over 8 q^4}
-{81\over 8 q^5}-{89\over 16 q^6} +O({1\over q^7})\big) \cr
c(q)~&=~ {1 \over \sqrt{\pi}}\big(
1-{6\over q}-{28\over q^2}+ {92\over q^3}+{196\over q^4}
+{224\over q^5}-{2412\over q^6}+O({1\over q^7})\big)  \cr}}}
The exponent $\alpha(q)$ is also found to tend to the Catalan value $3/2$
when $q$ tends to infinity, but appears to be a non-constant function
of $q$ for all finite $q$.
Remarkably, the first terms of the large $q$ series expansions of $R(q)$
and those of the correction to semi-meanders ${\bar R}_1(q)$ coincide!

In conclusion, all our results conspire to suggest that 
the semi-meanders undergo at $q=q_c$
a transition from a ``meander''-like regime $q<q_c$, governed by the 
meander partition function per bridge $R(q)\equiv {\bar R}_1(q)$, 
to another regime $q>q_c$, governed by ${\bar R}_0(q)$. 
The order parameter for this transition is clearly 
\eqn\orpam{ \lim_{n \to \infty} {1 \over n} \langle w \rangle_n~=~
\left\{\matrix{ \lambda(q) \  &{\rm for} \ \ q>q_c \cr
0 \ &{\rm for} \ \ q<q_c \cr} \right. }
which vanishes for $q<q_c$ (irrelevant winding, i.e. $\nu(q)<1$) and
is nonzero for $q>q_c$ (relevant winding, i.e. $\nu(q)=1$).
With the order parameter \orpam, the transition is found to be continuous,
as the leading coefficient $\lambda(q)$ \lamu\ vanishes at $q=q_c$. 
The smooth character of the transition is also visible from
the fact that ${\bar R}(q)$ and $R(q)$ approach each 
other tangentially at $q=q_c$, and that the coefficient 
${\bar c}_0(q)={\bar R}(q) \sigma(q)/q$ of the large-$q$ 
dominant contribution to ${\bar m}_n(q)$ vanishes at $q=q_c$ 
(see Fig.\plolamsig).

\newsec{Small $q$ behavior of the semi-meander polynomial}

The very existence of asymptotics of the form \asympto\ for the semi-meander
numbers, with a smooth enough function ${\bar R}(q)$ has highly 
non-trivial consequences
on the numbers ${\bar M}_n^{(k)}$.  We have already seen how the numbers
${\bar M}_n^{(n-k)}$, for large $n$ and finite $k$, are linked to each
other \pmnk\ in order for \asympto\ to hold for large $q$. Let us now
examine its consequences
on the small $q$ and large $n$ behavior of ${\bar m}_n(q)$.
Let us expand ${\bar m}_n(q)$ around $q=0$ up to order $k$, and take
the large $n$ asymptotics of each term in the expansion:
\eqn\expansion{\eqalign{
{\bar m}_n(q)~&=~\sum_{j=1}^k {\bar M}_n^{(j)} q^j + O(q^{k+1})\cr
&\sim {\bar c}(q)~{{\bar R}(q)^n \over n^{\gamma(q)}} \cr
&=~ {{\bar c} q \over n^{\gamma(q)}} R(0)^n 
\bigg(1+q {{\bar R}'(0) \over {\bar R}(0)}+O(q^2)\bigg)^n \cr
&=~ {\bar c}\, q {{\bar R}(0)^n\over n^\gamma}
\sum_{j=1}^k {n^{j-1} \over (j-1)!}  
\bigg(q {{\bar R}'(0) \over {\bar R}(0)}\bigg)^{j-1} +O(q^{k+1})  \cr }}
where we have only retained the leading $n$ asymptotics in each $q^j$ term,
and used the $q\to 0$ limits \recoq\ (actually, we have assumed that 
$q \sim 1/n$).
Comparing with the ${\bar M}_n$ asymptotics \actiga, we finally get
\eqn\ratiosm{ {{\bar M}_n^{(k)} \over {\bar M}_n^{(1)}}~\sim~ {1 \over (k-1)!}
\left(  n {{\bar R}'(0) \over {\bar R}(0)} \right)^{k-1} }
valid for large $n$ and finite $k$.  
This is actually very similar to the behavior of the 
${\bar M}_n^{(n-k)}\sim n^k/k!$ for large $n$ and finite $k$, 
and may be deduced from the main recursion relation for semi-meanders as well.
Indeed, the number ${\bar M}_n^{(k)}$ of semi-meanders of order $n$
with $k$ connected components is obtained from that of order $1$ (root)
by $(k-1)$ applications of the process (II) (see Fig.\treeofsm), and
$(n-k)$ applications of the process (I), whereas the ${\bar M}_n^{(1)}$
connected semi-meanders are obtained through the process (I) only.
Apart from the relative combinatorial factor ${k-1 \choose n-1}\sim
n^{k-1}/(k-1)!$ accounting for the $(k-1)$ choices of process (II)
among the total of $(n-1)$ steps, we must consider that whenever a step
(II) is chosen instead of a step (I),
some freedom in the overall choice is lost. Eq.\ratiosm\ tells us that
this corresponds to an average factor of
${\bar R}'(0)/{\bar R}(0)$ per step (II) taken instead 
of a step (I). We checked \ratiosm\ numerically, by 
performing a large $n$ extrapolation of the appropriate
ratio for a few values of $k$.
We find a very good agreement with the estimate
\eqn\estiratio{ {{\bar R}'(0) \over {\bar R}(0)} ~ \sim ~ 0.154(1) }
 
\newsec{Conclusion} 

In this paper, we have analyzed the meander problem in the language of
critical phenomena, by analogy with Self-Avoiding Walks. In particular,
we have displayed various scaling behaviors, involving both scaling
exponents and scaling functions.  We have presented strong evidence for
the existence of a phase transition for semi-meanders weighed by a
factor $q$ per connected component (road).  In a large-$q$ regime
($q>q_c$), the winding is found to be relevant, with a winding exponent
$\nu(q)=1$, while the configuration exponent $\gamma(q)=0$. Throughout
this phase, a finite fraction $\sigma(q)/\lambda(q)$ of the winding is
due to circles, i.e. circular roads with only one bridge, winding
around the source of the river.  In this regime, the partition function
per bridge for semi-meanders ${\bar R}(q)$ is strictly larger than that
of meanders $R(q)$.  The particular form of its large $q$ series
expansion in $1/q$ \resultasympto\ with slowly alternating integer
coefficients, which furthermore grow very slowly with the order, and
its purported re-summation (D.14), suggest a possible re-expressionion
in terms of modular forms of $q$, yet to be found.  In a low-$q$ regime
$q<q_c$, ${\bar R}(q)$ and $R(q)$ coincide, in agreement with an
irrelevant winding $\nu(q)<1$.  The exponent $\gamma(q)$ is no longer
$0$, but a strictly positive function of $q$.

We have estimated the value of the transition point $q_c \simeq 2$
with an excellent precision, and we conjecture that $q_c=2$
exactly.  This special value of $q$ has already been singled
out in the algebraic study of the meander problem, in connection
with the Temperley-Lieb algebra \NTLA. There, we have been
able to re-express the meander and semi-meander partition functions 
as that of some Restricted Solid-On-Solid model, whose 
Boltzmann weights are positive precisely iff $q\geq 2$, 
indicating very different behaviors for $q<2$ and $q>2$.

In the large-$q$ phase, in addition to the exact values $\nu(q)=1$ 
and $\gamma(q)=0$, we can  use the asymptotic expansion \resultasympto\
to get ${\bar R}(q)$ with a very good precision.  
The somewhat sub-leading meander quantities $R(q)$ and $\alpha(q)$ are
more difficult to evaluate in this regime.

A number of questions remain unsettled.  
There still remains to find the varying exponents 
$\gamma(q)$ and $\nu(q)$ in the $q<2$ regime, as well as 
the precise value of $R(q)={\bar R}(q)$. 
Although we improved our numerical estimates, we are limited
to conjectures. For $q=0$, we confirm a previous conjecture 
\DGG\ that $\gamma=2$, and that \LZ\ $\alpha=7/2$.
We also conclude from the numerical analysis that $\nu(0)\simeq 0.52(1)$
is definitely not equal to the trivial random-walk exponent $1/2$.
For $q=2$, we have an amusing guess for 
$R(2)={\bar R}(2)=\pi \sqrt{2}=\Gamma(1/4)\Gamma(3/4)=4.442...$
inspired from possible infinite product formulas for ${\bar R}(q)$.

\noindent{\bf Acknowledgments}

We thank M. Bauer for a critical reading of the manuscript.

\appendix{A}{Algorithm for the enumeration of semi-meanders}

The following is a simple computer algorithm directly inspired by the
recursion relation of Sect.3.1.

There, it has been shown that one can construct a {\it tree} of all the
semi-meanders generated recursively with processes (I) and (II), as
displayed in Fig.\treeofsm.  
Each {\it node} at {\it depth} $n$ represents a semi-meander of order $n$.
To have a finite cost of computation, the order is limited to $nmax$ and
the nodes of depth $nmax$ appear to be {\it leaves}. 
The algorithm we
used consists in visiting all the leaves from left to right, following
the branches, as a (clever) squirrel.

To do that, the rules are :
\item{(a)} The squirrel starts at the root (upper node).
\item{(b)} When the squirrel is on a intermediate node (not a leaf), he
follows the leftmost branch which it has not yet visited and the depth
increases by 1.   
If all the branches of a given node have been visited, 
the squirrel goes back up
one level and the depth decreases by 1.
\item{(c)} When the squirrel is on a leaf (depth $nmax$), it goes back up
one level and the depth becomes $(nmax -1)$.

The reader can convince himself that the above rules describe a systematic
and complete visit of the tree.  Of course, when the squirrel is on a node,
it can measure a lot of interesting quantities like the number of connected
components, the winding number...
These measures are added up and analyzed at the end of the enumeration.

{}From a Fortran point of view, many representations of (semi-)meanders are
possible.  In the open-river formulation, each semi-meander is made of a 
lower rainbow arch configuration
(which we need not code) and an upper arch configuration
of order $n$.  
For convenience, we label the $2n$ bridges of river from $i=-n+1$ to $i=n$, 
and the system of arches is described
by a sequence of integers
$\{A(i); i=-n+1,...,n\}$, where $A(i)\in \{-n+1,...,n\}$ is the label
of the bridge connected with the bridge $i$.
The following Fortran program enumerates the connected semi-meanders.  
For simplicity, only the process (I) is coded and the number of connected
components is always $k=1$.
The arch to be broken $(j,A(j))$ begins at the bridge
$j$ and ends at the bridge $A(j)$. The process (I) splits this arch into
two arches $(-n,j)$ and $(A(j),n+1)$. When the squirrel climbs back up
one level, the two extremal arches $(-n+1,A(-n+1))$ and $(A(n),n)$ 
are re-sealed to give one arch $(A(-n+1),A(n))$.  At this stage,
we know that the next arch to break starts at bridge $j=A(n)+1$.
This ensures the completeness of the algorithm.

\def\tvi{\vrule height 10pt depth 6pt width 0pt}
\def\tv{\tvi\vrule}

$$\encadre{\vbox{
\offinterlineskip
\halign{ \quad \hfill # &  # \hfill &  #  \hfill &  #  \hfill \cr
&\ &PARAMETER (nmax = 14) &! maximal order \cr
&\ &INTEGER A(-nmax+1:nmax) &! arch representation \cr
&\ &INTEGER Sm(nmax) &! semi-meander counter \cr
&\ &INTEGER n &! current depth (or order) \cr
&\ &INTEGER j &! next branch to visit \cr
&\ &DATA  n, Sm /0,  nmax*0/ &! n and Sm initialized to 0 \cr
&\ &\ &\ \cr
&\ &A(0) =  1 &! single-arch semi-meander  \cr
&\ &A(1) =  0 &\ \cr
&\ &\ &\ \cr
&2 &n  =  n + 1 &! a new node is visited \cr
&\ &Sm(n)  =  Sm(n) + 1 &\ \cr
&\ &j  =  -n + 1 &! leftmost (exterior) arch  \cr
&1 &IF((n.EQ.nmax).OR.(j.EQ.n+1)) GOTO 3 &! up or down ? \cr
&\ &\ &\ \cr
&\ &A(A(j))  =  n+1 &! go down with process (I) \cr
&\ &A(n+1)  =  A(j) &\ \cr
&\ &A(j)  =  -n &\ \cr
&\ &A(-n)  =  j &\ \cr
&\ &GOTO  2 &\ \cr
&\ &\ &\ \cr
&3 &A(A(-n+1))  =  A(n) &! going up \cr
&\ &A(A(n))  =  A(-n+1) &\ \cr
&\ &j  =  A(n)+1 &! next arch to break \cr
&\ &n = n - 1 &\ \cr
&\ &IF (n .GT. 1) GOTO 1 &\ \cr
&\ &\ &\ \cr
&\ &PRINT '(i3, i15)', (n, Sm(n), n = 1, nmax) &\ \cr
&\ &END \cr
}}}$$

It is possible to use the left-right symmetry to divide the work by two.
It is also possible to adapt the program for a parallel computer.  For
that, an intermediate size ($n1 = 11$, for instance) is chosen.
A first (little) run is made with $nmax = n1$, which gives $\bar M(n1)$
leaves.  In a second (big and parallelized) run, each of these leaves is
now taken as the root of a (sub-)tree and treated independently of the
others.  At the end, all the results of the sub-trees are collected.
The calculations of this article have be done on the parallel Cray-T3D
(128 processors) of the CEA-Grenoble, with approximately 7500 hours 
$\times$ processors.

We have computed $\bar M_n$ (the (pure) semi-meander number of order
$n$) up to $n=29$, $\bar M_n^{(k)}$ (semi-meander number of order $n$
with $k$ connected components) up to $n=27$ and the other quantities up
to $n=24$.  The reader can obtain an electronic copy of the 
numerical data upon request to the authors.  We content ourselves
with giving, on Table I below, the $\bar M_n$ and our last 
row ($n=27$) of the ${\bar M}_n^{(k)}$.

$$\vbox{\font\bidon=cmr8 \bidon
\offinterlineskip
\halign{\tv \quad \hfill # & \hfill \ # \tv 
& \quad \hfill # &  \hfill # \tv \tv 
& \quad \hfill # &  \hfill # \tv 
& \quad \hfill # &  \hfill # \tv \cr
\noalign{\hrule}
$n$ & $\bar M_n$ & $n$ & $\bar M_n$ & $k$ & $\bar M_{27}^{(k)}$ 
& $k$ & $\bar M_{27}^{(k)}$\cr
\noalign{\hrule}
1  &  1  & 16  &  1053874  &  1  &  369192702554  & 16  &  2376167414\cr
2  &  1  & 17  &  3328188  &   2  &  2266436498400 & 17  &  628492938\cr
3  &  2  & 18  &  10274466  &   3  &  6454265995454 &  18  &  153966062\cr
4  &  4  & 19  &  32786630  &   4  &  11409453277272 &  19  &  34735627\cr
5  &  10  & 20  &  102511418  &  5  &  14161346139866  &  20  &  7159268\cr
6  &  24  & 21  &  329903058  &  6  &  13266154255196 &  21  &  1333214\cr
7  &  66  & 22  &  1042277722  &  7  &  9870806627980 &  22  &  220892\cr
8  &  174  & 23  &  3377919260  &  8  &  6074897248976 &  23  &  31851\cr
9  &  504  & 24  &  10765024432  &  9  &  3199508682588 &  24  &  3866\cr
10  &  1406  & 25  &  35095839848  &  10  &  1483533803900 &  25  &  374\cr
11  &  4210  & 26  &  112670468128  &  11  &  619231827340  &  26  &  26\cr
12  &  12198  & 27  &  369192702554  &  12  &  236416286832 &  27  &  1\cr
13  &  37378  & 28  &  1192724674590  &  13  &  83407238044 &  &  \cr
14  &  111278  & 29  &  3925446804750  &  14  &  27346198448 &  &  \cr
15  &  346846  &     &                 &  15  &  8352021621 &  &  \cr
\noalign{\hrule}
}}$$
\noindent{\bf Table I:} The numbers $\bar M_n^{(k)}$ of semi-meanders
of order $n$ with $k$ connected components, obtained by exact
enumeration on the computer:  on the left, the one-component
semi-meander numbers ($k=1$) are given for $n \le 29$; on the right, 
$n$ is fixed to 27 and $1 \leq k \leq n$.  
The $\bar M_n^{(k)}$ for $n < 27$ can be obtained by request to the authors.

\appendix{B}{Correction terms for semi-meanders with large 
number of connected components}

In this appendix, we show that the corrections to the polynomial 
expression \asypolsm\ for the numbers ${\bar M}_n^{(n-k)}$ 
occur only for $n \leq 2k-2$, and
derive the first correction $\mu_0^{(k)}$ of \asypolsm\ 
for $n=2k-2$.
The result reads 
\eqn\resucoral{ \mu_0^{(k)}~=~ c_{k-1} }
where the $c_n$ are the Catalan numbers \catal.

As explained in Sect.4, the polynomial part of ${\bar M}_n^{(n-k)}$ is
generically obtained as a sum of combinatorial factors, counting all
the possible occurrences of perturbations of the leading semi-meander
(order $n$, $n$ components) which have the same order $n$, but have
only $n-k$ connected components.  A perturbation, by definition,  is
made of a core,
which does not contain any circle,
supplemented by circles. 
The core is generically made of $p$ irreducible semi-meanders
with a total order $n_0$, and a total of $(n_0-k)$ connected components.
By irreducible, we mean that no circle can separate the
semi-meander in two disconnected pieces.
This core is completed by $(n -n_0)$ circles,
to form a semi-meander of order $n$. 
Enumerating all perturbations of the leading meander with $n-k$
components amounts to enumerating all the ways of completing cores by
circles.  There are exactly
\eqn\cornumb{ {n-n_0+p\choose p}~=~
{(n-n_0+p)(n-n_0+p-1)...(n-n_0+1) \over p!} }
possible decorations of the above core by circles. The polynomial form 
\cornumb\ of the combinatorial factor is however only valid
for $n \geq n_0-p$. When $n\leq n_0-p-1$, we have to add a correction 
to the polynomial form to get a vanishing result, 
equal to $(-1)^{p+1}$ for the largest $n=n_0-p-1$. 
The largest $n$ at which such corrections occur is obtained by
maximizing $n_0-p$.
As the core does not contain any circle, each of its $n_0-k$ 
connected components is at least of order $2$, hence
\eqn\ineqdif{ n_0 ~\geq~ 2 (n_0-k)\quad \Rightarrow \quad n_0\leq 2k }
This inequality is saturated for a core made of $k$ kidneys. Minimizing
$p$ consists in taking $(k-1)$ arbitrary kidneys included in one,
for which $p=1$.
We therefore find that the first correction to the polynomial behavior 
\asypolsm\ occurs at $n=2k-2$.

\fig{Arbitrarily disjoint or included kidneys are equivalent
to symmetric meanders upon folding back the river, as indicated 
by the arrows. The latter are in one-to-one correspondence with
arch configurations of the same order.}{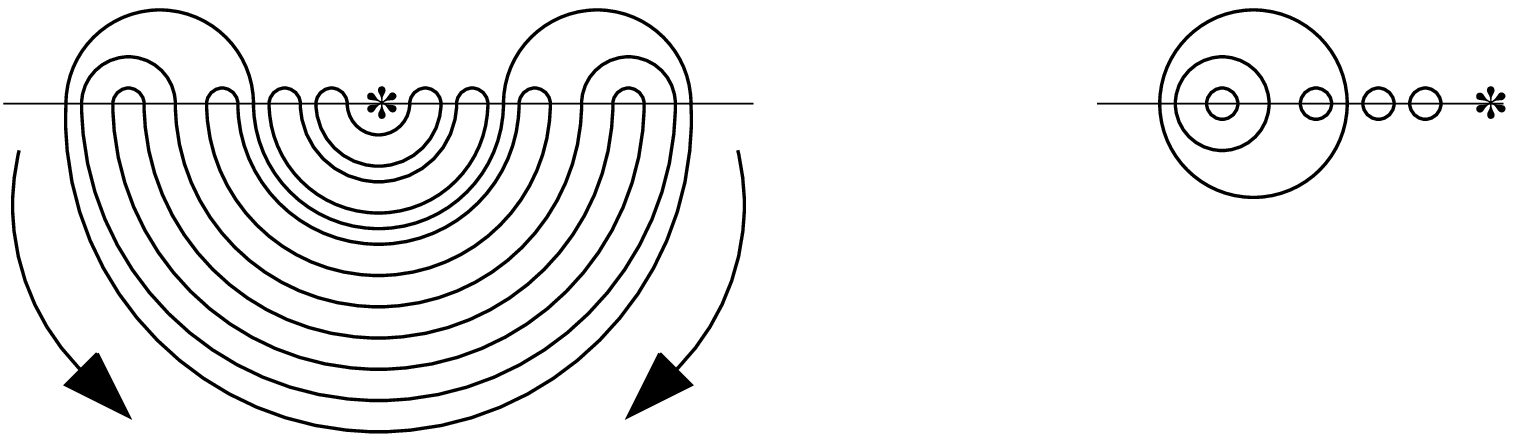}{10.cm}
\figlabel\kidfold

There are exactly $c_{k-1}$ possible choices of these $(k-1)$ kidneys,
in one-to-one correspondence with arch configurations of order $(k-1)$
as illustrated in
Fig.\kidfold,
upon a folding procedure. 
Each corresponding core contributes $1$ to the
correction $\mu_0^{(k)}$, 
which completes the proof of \resucoral.

More generally, we expect corrections to the polynomial part of
${\bar M}_n^{(n-k)}$ for all $k \leq n \leq 2k-2$, hence the form \asypolsm.
In appendix C, the structure of the first $9$ successive corrections is found,
together with the polynomial part of ${\bar M}_n^{(n-k)}$, up to $k=18$.

\appendix{C}{Fine structure of the semi-meander numbers}

As shown in appendix B, the polynomial part in \asypolsm\ of the
semi-meander numbers
$p_k(n)$  is equal to the semi-meander number 
${\bar M}_n^{(n-k)}$ for $n \geq 2k-1$.
In addition, the polynomials $p_k$ are subject to the quasi-recursion
relation \constrpol, which leaves only two coefficients of $p_k$ to be
fixed, once the $p_l$, $l \leq k-1$ are known (we have $p_0(n)=1$).  It
is therefore straightforward to
derive the first $13$ polynomials $p_k$ from our numerical data, as only
the two values $p_k(2k-1)={\bar M}_{2k-1}^{(k-1)}$ and
$p_k(2k)={\bar M}_{2k}^{(k)}$ are required, and the corresponding 
numbers are known up to $k=13$.

However, knowing the $p_k$'s up to $k=13$, we get a list of all the
corrections
\eqn\correctpol{ \mu_j^{(k)}~=~ {\bar M}_{2k-2-j}^{(k-2-j)} - p_k(2k-2-j) } 
to these polynomials needed to get the correct values of 
${\bar M}_n^{(n-k)}$, up to $k=13$. 
The resulting table of corrections displays a remarkable structure
(very close to that of meander numbers, studied in appendix E), 
which reads as follows, according to the parity of $j$
\eqn\corectrue{\eqalign{
{\bar \mu}_{2j}^{(p+1+j)}~&=~c_p {2^j \over j!} 
{(2p+1)! \over (2p+3j)!} \Pi_{4j-1}(n) \quad {\rm with} \, n=2p \cr
{\bar \mu}_{2j+1}^{(p+1+j)}~&=~c_p 3 {2^j \over j!} {(2p)! \over (2p+3j)!}
\Pi_{4j}(n)\quad {\rm with} \, n=2p-1 \cr}}
valid for all $j \geq 0$, where $c_p$ are the Catalan numbers \catal,
and the $\Pi_k(n)$ are monic polynomials of degree $k$. 

Proceeding in parallel (determining the $p_k$ and the corrections
simultaneously), we can proceed further up to $k=18$, with the result
$$\eqalign{
p_0(x)~&=~1 \cr
p_1(x)~&=~x-1\cr
p_2(x)~&=~(x^2 +    x   -   8 )/2 \cr
p_3(x)~&=~(x^3 +  6 x^2 -  31 x   -   24 )/3! \cr
p_4(x)~&=~(x^4 + 14 x^3 -  49 x^2 -  254 x )/4! \cr
p_5(x)~&=~(x^5 + 25 x^4 -  15 x^3 - 1105 x^2 - 1066 x   
+ 1680)/5! \cr
p_6(x)~&=~(x^6 + 39 x^5 + 145 x^4 - 2895 x^3 -10226 x^2 
+ 8616 x  +31680)/6!\cr
p_7(x)~&=~(x^7+  56 x^6 + 532 x^5 - 5110 x^4 -50141 x^3 
-20146 x^2+377208 x\cr
&+282240)/7!\cr
p_8(x)~&=~(x^8+76 x^7 + 1274 x^6 -5264 x^5-165991 x^4
-422156 x^3+1979116 x^2\cr
&+6031824 x +1128960)/8!\cr
p_9(x)~&=~(x^9+99 x^8 +2526 x^7+2646 x^6-413511 x^5
-2570589 x^4+4826744 x^3\cr
&+55185444 x^2+54007920 x-29756160)/9!\cr
p_{10}(x)~&=~(x^{10}+125 x^9+4470 x^8+30090 x^7
-803607 x^6-10282755 x^5\cr
&-6206320 x^4+302065660 x^3
+838000656 x^2-179320320 x-914457600)/10!\cr
p_{11}(x)~&=~(x^{11}+154 x^{10}+7315 x^9+96360 x^8
-1170477 x^7-31531038 x^6\cr
&-116748115 x^5+1085347340 x^4
+7183991276 x^3+4813856784 x^2\cr
&-20917209600 x
-15487718400)/11!\cr
p_{12}(x)~&=~(x^{12}+186 x^{11}+11297 x^{10}+231330 x^9
-921657 x^8-78859242 x^7\cr
&-632084629 x^6+2301195270 x^5
+41279402956 x^4+93554770056 x^3\cr
&-181951879968 x^2
-528315782400 x-281652940800)/12!\cr
p_{13}(x)~&=~(x^{13}+221 x^{12}+16679 x^{11}+478621 x^{10}
+1380093 x^9-164767317 x^8\cr
&-2382680443 x^7-496896257 x^6
+172979664286 x^5+862378655996 x^4\cr
&-433580004936 x^3
-7916932037664 x^2 -9843473134080 x\cr
&-4807260057600)/13!\cr}$$
\vfill\eject
\eqn\resultap{\eqalign{
p_{14}(x)~&=~(x^{14}+259 x^{13}+23751 x^{12}+899171 x^{11}
+8661653 x^{10}\cr
&-285047763 x^9-7202313547 x^8-30034731727 x^7
+538444152246 x^6\cr
&+5357996127484 x^5+6470339335096 x^4
-67697579511744 x^3\cr
&-191470854038400 x^2 
-156970678394880 x -54050540544000)/14!\cr
p_{15}(x)~&=~(x^{15}+300 x^{14}+32830 x^{13}+1575210 x^{12}
+26278252 x^{11}\cr
&-378408030 x^{10}-18419182210 x^9
-164903537370 x^8+1145711810243 x^7\cr
&+25071998561610 x^6
+89293562501780 x^5-342807873156840 x^4\cr
&-2342105695034496 x^3
-3394586375786880 x^2-1602031548902400 x\cr
&+1953665505792000)/15!\cr 
p_{16}(x)~&=~(x^{16}+344 x^{15}+44260 x^{14}+2614640 x^{13}
+63321622 x^{12}\cr
&-244465312 x^{11}-40783574260 x^{10}
-621949980080 x^9+743774155553 x^8\cr
&+92574072382792 x^7
+650521234967240 x^6-658128229828160 x^5\cr
&-19409328228712176 x^4
-53224642530877824 x^3-35515613029674240 x^2
\cr
&+48894886046361600 x+121101107871744000)/16!\cr
p_{17}(x)~&=~(x^{17}+391 x^{16}+58412 x^{15}+4155820 x^{14}
+134381702 x^{13}\cr
&+605876362 x^{12}-78440970036 x^{11}
-1905575960860 x^{10}\cr
&-7182404321807 x^9+271696014206903 x^8
+3438145428617536 x^7\cr
&+5440767496706360 x^6
-113308202712264496 x^5-594837570662080656 x^4\cr
&-742844861844630912 x^3
+780842535860839680 x^2+4213612687918233600 x\cr
&+2696110704967680000)/17! \cr
p_{18}(x)~&=~
(x^{18}+441 x^{17}+75684 x^{16}+6372756 x^{15}
+261854502 x^{14}\cr
&+3203791542 x^{13}-128136377252 x^{12}
-5023361538468 x^{11}\cr
&-45509495478447 x^{10}
+602361827803593 x^9+14435051961445752 x^8\cr
&+67015410007677768 x^7-451434378729887216 x^6
-4879430147272561776 x^5\cr
&-11678841781394909184 x^4+6041803064009266944 x^3 \cr
&+78098750156766044160 x^2
+148841653993843507200 x\cr
&-4417637856952320000)/18!\cr}}

The first $\Pi$'s read,
\eqn\firstpis{\eqalign{
\Pi_3(x)~&=~x^3 +{33 \over 2} x^2+{49 \over 2} x+12\cr
\Pi_7(x)~&=~x^7+44 x^6+{1397 \over 2} x^5+3371 x^4+{13229\over 2} 
x^3+11735 x^2+6336 x+1440\cr
\Pi_{11}(x)~&=~x^{11}+{163 \over 2} x^{10}+2814 x^9+{198453\over 4} 
x^8+405471 x^7+1586409 x^6\cr
&+4522625 x^5+{53075437\over 4} x^4+13577913 x^3
+14321997 x^2+3974616 x\cr
\Pi_{15}(x)~&=~x^{15}+129 x^{14}+7462 x^{13}+248370 x^{12}+
{9911591\over 2} x^{11}+{11187899 \over 2} x^{10}\cr
&+345882086 x^9+1392155925 x^8+
{12479885017 \over 2} x^7+{44260908189 \over 2} x^6\cr
&+30440330807 x^5+71900294130 x^4
+17319985020 x^3\cr
&-28987682544 x^2-36059230080 x-10059033600\cr
\Pi_0(x)~&=~1\cr
\Pi_4(x)~&=~x^4+{47 \over 3}x^3 +31 x^2+{37 \over 3} x+20\cr
\Pi_8(x)~&=~x^8+{124 \over 3} x^7+{1283 \over 2} x^6+
{18479\over 6} x^5+{8543\over 2} x^4+{72077\over 6} 
x^3\cr
&+26106 x^2-4994 x-840\cr
\Pi_{12}(x)~&=~x^{12}+77 x^{11}+{5135 \over 2} x^{10}+
{88129\over 2} x^9+341193 
x^8+1023192 x^7\cr
&+{ 5147959\over 2} x^6+{35955101\over 2} x^5+
23942231 x^4-16806044 x^3\cr
&+18500028 x^2-21290040 x-6350400\cr
\Pi_{16}(x)~&=~x^{16}+{368 \over 3} x^{15}+6845 x^{14}+
{664235\over 3} x^{13}+{8557681\over 2} x^{12}+
{135562460\over 3} x^{11}\cr
&+{452366205 \over 2} x^{10}+
{1940546440\over 3} x^9+{11216097197\over 2} x^8+
{89505296956 \over 3} x^7\cr
&+{9305821085 \over 2} x^6+
{27505324345\over 3} x^5+164921583110 x^4
-422241984828 x^3\cr
&-328334271840 x^2-252250683840 x-40475635200\cr}}

\appendix{D}{Semi-meanders and winding}

In this appendix, we study the numbers ${\bar M}_n^{(k)}(w)$ of
semi-meanders of order $n$, with $k$ connected components
and winding $w$. The winding $w$ has the same parity as
the order $n$, hence we consider the following
generating function
\eqn\genwk{ {\bar m}_n(q,t)~=~\sum_{k=0}^{n-1} q^{n-k} 
\sum_{j=0}^{{\rm min}(k,[n/2])} t^{j} {\bar M}_n^{(n-k)}(n-2j) }
interpreted as the partition function for semi-meanders weighed by
$q$ per connected component and $1/\sqrt{t}$ per winding unit 
(up to a global normalization factor). In the large $q$ limit
we now repeat the analysis performed in sect.4.1, by 
Taylor-expanding ${\rm Log}({\bar m}_n(q,t)/q^n)$
order by order in $1/q$, in the same way as we did previously for
${\rm Log}({\bar m}_n(q)/q^n)$.  This also relies on the 
identification of the numbers ${\bar  M}_n^{(n-k)}(n-2j)$ as 
polynomials of $n$, with special re-exponentiation properties, 
as discussed now.

For $k=0$, the corresponding leading semi-meander
has winding $w=n$, hence
\eqn\mbwin{{\bar M}_n^{(n)}(n-2j)~=~\delta_{j,0} }
For large $n$ and finite $k\geq 1$, the winding of a
semi-meander of order $n$ with $n-k$ connected components
may only take the values $(n-2k)$, $(n-2k+2)$, ...,$(n-2)$.
Indeed, if $C$ denotes the number of circles of such a semi-meander 
with winding $(n-2j)$, there
are $(n-k-C)$ connected components which are not circles, 
hence which occupy two or more bridges. Therefore we have the following lower 
bound on the order of the semi-meander
\eqn\lobounor{ n ~\geq~ C+2(n-k-C)\ \  \Leftrightarrow \ \ C~\geq ~n-2k}
and therefore, as each circle contributes $1$ to the total winding,
\eqn\winlowb{ w ~\geq ~C \geq n-2k}

When $k=1$, all the one-kidney perturbations \kidneyper\
of the leading semi-meander have winding $w=(n-2)$, hence 
\eqn\mbwinone{{\bar M}_n^{(n-1)}(n-2)~=~n-1}

When $k=2$, let us reexamine the various semi-meanders obtained 
in Fig.\firstorder: the perturbations of Figs.\firstorder\ $(c)_{1,2}$ have
winding $(n-4)$, whereas those of Figs.\firstorder\ $(c)_{3,4}$ have
winding $(n-2)$. Hence we have
\eqn\winthr{\eqalign{ {\bar M}_n^{(n-2)}(n-4)~&=~
{n(n-3)\over 2}+\delta_{n,2} \cr
{\bar M}_n^{(n-2)}(n-2)~&=~2(n-2)\cr} }
The correction is ad hoc to yield a zero answer when $n=2$.

With a little more patience, the enumeration of the semi-meanders of 
order $n$ with $(n-3)$ connected components and fixed winding yields
\eqn\winfour{\eqalign{ 
{\bar M}_n^{(n-3)}(n-6)~&=~{n(n-1)(n-5)\over 6}
+2 \delta_{n,3}+2 \delta_{n,4}\cr
{\bar M}_n^{(n-3)}(n-4)~&=~2 (n^2-4n+1)+4 \delta_{n,3}\cr
{\bar M}_n^{(n-3)}(n-2)~&=~2 (n-3) \cr}}

In general, the ${\bar M}_n^{(n-k)}(w)$ form a decomposition of the 
${\bar M}_n^{(n-k)}$, in the sense that
\eqn\filter{ \sum_w {\bar M}_n^{(n-k)}(w)~=~ {\bar M}_n^{(n-k)}}
In a way similar to the ${\bar M}_n^{(n-k)}$, we expect the
numbers ${\bar M}_n^{(n-k)}(n-2j)$ to be, for large enough $n$, 
some polynomials of $n$, whose coefficients depend only on $j$ and $k$.
For small $n$, some corrections have to be added to recover the actual numbers
{}from their polynomial part.
More precisely, one can show that
\eqn\polwind{  {\bar M}_n^{(n-k)}(n-2j)~=~ p_j^{(k)}(n)+
\sum_{m=0}^{j-2} \delta_{n,k+j-m-2}\, \eta_{m}^{(k,j)} }
for $1\leq j\leq k\leq n$, 
where $p_j^{(k)}(x)$ is a polynomial of degree $j$ of $x$
($p_0^{(0)}(x)=1$), whose coefficients 
depend on $j$ and $k$, and the $\eta_m^{(j,k)}$ are non-negative 
integer corrections.  
In particular, according to \filter, we must have
\eqn\fifil{ \sum_{j=1}^k p_j^{(k)}(x) ~=~ p_k(x) }
We have computed the polynomials $p_j^{(k)}(x)$ for 
$0 \leq j \leq k\leq 14$, by using exact enumeration data
on the ${\bar M}_n^{(n-k)}(n-2j)$ for $0\leq j\leq k \leq n \leq 24$,
and the re-exponentiation trick described in sect.4.1.
These lead to the large $n$ asymptotics of the partition function \genwk\
in the large $q$ regime
\eqn\asymwk{ {\bar m}_n(q,t)~\sim~ {\bar c}(q,t)\, 
{{\bar R}(q,t)^n \over n^{\gamma(q,t)} } }
with $\gamma(q,t)=0$ as before, and with the following 
large $q$ series expansions
\eqn\rofwk{
\eqalign{{\bar R}(q,t)~&=~q+t+{2t \over q}+{2t\over q^2}+{2t\over q^3}
+{2t(t-1)^2\over q^4}+{2t(t^2-4t+1)\over q^5}\cr
&+{2t(3t^2-8t+1)\over q^6}+{2t(4t^4-8t^3+9t^2-12t+1)\over q^7}\cr
&+{2t(10t^4-26t^3+28t^2-18t+1)\over q^8}
+{2t(6t^5+12t^4-61t^3+64t^2-24t+1)\over q^9}\cr
&+{2t(24t^6-52t^5+71t^4-137t^3+131t^2-32t+1)\over q^{10}}\cr
&+{2t(101t^6-260t^5+308t^4-324t^3+237t^2-40t+1)\over q^{11}}\cr
&+{2t(90t^7+50t^6-610t^5+894t^4-726t^3+400t^2-50t+1)\over q^{12}}\cr
&+{2t(173t^8-243t^7+409t^6-1507t^5+2237t^4-1564t^3+631t^2-60t+1)
\over q^{13}}\cr
&+O({1\over q^{14}}) \cr }}
\eqn\cofwk{
\eqalign{{\bar c}(q,t)~&=~1-{t\over q}
-{4t\over q^2}+{2t(t-3)\over q^3}+{8t(t-1)\over q^4}
-{2t(7t^2-19t+5)\over q^5}\cr
&+{2t(2t^3-13t^2+39t-6)\over q^6}+{2t(18t^3-62t^2+79t-7)\over q^7}\cr
&-{2t(45t^4-122t^3+183t^2-128t+8)\over q^8}\cr
&+{2t(9t^5-138t^4+406t^3-514t^2+205t-9)\over q^9}\cr
&+{2t(69t^5-475t^4+1138t^3-1143t^2+295t-10)\over q^{10}}\cr
&-{2t(353t^6-1183t^5+2104t^4-3012t^3+2342t^2-421t+11)\over q^{11}}\cr
&+{2t(56t^7-1553t^6+4986t^5-7692t^4+7668t^3-4311t^2+564t-12)\over q^{12}}\cr
&-{2t(230t^7+3888t^6-14710t^5+22124t^4-17898t^3+7490t^2-751t+13)
\over q^{13}}\cr
&-{2t\over q^{14}}(3123t^8-10053t^7+22655t^6-45694t^5+58185t^4-39176t^3\cr
&+12215t^2-959t+14)+O({1\over q^{15}})\cr}}
The previous results 
\resultasympto\ are recovered up to order $14$ in $1/q$
by taking $t=1$. Let us look more 
closely at the expressions \rofwk\ and \cofwk\ above: we may Taylor-expand
them as functions of $t$ for small $t$. 
Remarkably, due to the structure of the coefficients 
of $p_j^{(k)}(x)$, which are themselves polynomials of $k$ for fixed $j$
(e.g. we have $p_1^{(k)}(x)=2(x-k)$ for all $k \geq 2$ while
$p_1^{(1)}(x)=x-1$), we have been able to re-sum the large $q$ series
coefficients of this expansion
up to order $3$ in $t$, in the following form
\eqn\rtaysum{
\eqalign{
{\bar R}(q,t)~&=q+t{q+1 \over q-1} -4 t^2 {1 \over (q-1)^2 (q^2-1)}\cr
&+4 t^3 {q^7-q^6+5q^4+11q^3+8q^2+q-1\over q^2(q-1)^2(q^2-1)^2(q^3-1)}
+O(t^4)\cr
{\bar c}(q,t)~&=~1-t{q^2+2q-1\over q(q-1)^2}
+2t^2{q^5+2q^4+10q^3+q^2-3q+1\over q^2(q-1)^2(q^2-1)^2}\cr
&-2{t^3\over q^4(q-1)(q^2-1)^3(q^3-1)^2}(7q^{12}+6q^{11}
+28q^{10}+89q^9+193q^8\cr
&+228q^7+147q^6
+35q^5-14q^4-5q^3+6q^2+q-1)
+O(t^4) \cr}}
This structure is very reminiscent of the large $Q$
$Q$-states Potts model free energy \BAX, and suggests a possible
expression in terms of modular forms of $q$.

The average winding in large $q$ semi-meanders is obtained by the formula
\eqn\larwinsm{ \langle w \rangle_n(q)~=~ 
n -2 t {d \over dt} {\rm Log}\, {\bar m}_n(q,t)\big\vert_{t=1} }
leading to the expression \asymwin\ through the identifications
\eqn\relamrb{
\eqalign{ \lambda(q)~&=~ 1-2 t{d \over dt} {\rm Log}\,{\bar R}(q,t)
\big\vert_{t=1}\cr
\mu(q)~&=~-2 t{d \over dt} {\rm Log}\,{\bar c}(q,t)
\big\vert_{t=1}\cr}}

\appendix{E}{Fine structure of the meander numbers}

By simple inspection of the meander numbers
(which we read from the semi-meander numbers with zero winding), 
we have found the 
following structure for the numbers $M_n^{(n-k)}$
\eqn\formea{M_n^{(n-k)}~=~ c_n ~ r_k(n)}
where $c_n$ is a Catalan number \catal ,
$r_k(x)$ is a rational fraction of $x$ with total degree $k$.
More precisely, we have, $r_0(n)=1$ and for $k \geq 1$
\eqn\reqlp{\eqalign{ r_k(n)~&=~{2^k  n! (n+1)! \over k! (n-k-1)! (n+2k)!} 
\varphi_{2k-2}(n)\cr
&=~ {2^k \over k!} 
{n(n-1)(n-2)...(n-k) \over (n+2)(n+3)...(n+2k)}\varphi_{2k-2}(n) \cr}}
where $\varphi_{2k-2}(x)$ are monic polynomials of $x$ of degree $(2k-2)$.
They read, for $k=1,2,...,6$
\eqn\ratmean{\eqalign{
\varphi_0(x)~&=~1 \cr
\varphi_2(x)~&=~x^2 +7x -2\cr
\varphi_4(x)~&=~x^4+20 x^3+107 x^2-107 x+15\cr
\varphi_6(x)~&=~x^6+39 x^5 +547 x^4 +2565 x^3- 5474 x^2+2382 x-672\cr
\varphi_8(x)~&=~x^8+64 x^7+1646 x^6+20074 x^5 
+83669 x^4-323444 x^3\cr
&+ 257134 x^2-155604 x+45360\cr
\varphi_{10}(x)~&=~x^{10}+95 x^9+3840 x^8+83070 x^7
+940443 x^6+3382455 x^5\cr
&-22294735 x^4+27662860 x^3-26147139 x^2
+16354530 x-4098600\cr}}

\listrefs

\bye